\documentclass[acmsmall,screen]{acmart}
\acmJournal{TOSEM}

\usepackage[utf8]{inputenc}
\usepackage[english]{babel}

\usepackage{graphicx}
\DeclareGraphicsExtensions{.pdf,.png,.jpg}

\usepackage{booktabs}
\usepackage{array}
\usepackage{enumitem}
\usepackage{xspace}
\usepackage[ruled,vlined,linesnumbered]{algorithm2e}

\usepackage{xcolor}
\usepackage[most]{tcolorbox}
\usepackage{tikz}
\usetikzlibrary{arrows.meta,backgrounds,fit}

\definecolor{codecomment}{RGB}{92,92,92}
\definecolor{codeframe}{RGB}{130,130,130}
\definecolor{codepanel}{RGB}{248,248,248}
\usepackage{listings}
\lstdefinestyle{mystyle}{
  commentstyle=\color{codecomment},
  keywordstyle=\bfseries,
  stringstyle=\color{black},
  basicstyle=\ttfamily\scriptsize,
  breaklines=true,
  captionpos=b,
  columns=fullflexible,
  keepspaces=true,
  showstringspaces=false,
  tabsize=2,
  aboveskip=0pt,
  belowskip=0pt
}
\tcbset{
  codepiece/.style={
    enhanced,
    colback=codepanel,
    colframe=codeframe,
    boxrule=0.35pt,
    arc=0.8mm,
    left=2mm,
    right=2mm,
    top=1.2mm,
    bottom=1.2mm,
    boxsep=0pt
  }
}

\usepackage[capitalize,nameinlink]{cleveref}

\newcommand{\denseitems}{%
  \setlength{\itemsep}{1pt}%
  \setlength{\parskip}{0pt}%
}

\newcommand{\papertitle}{CPyGraph: A Version-Aware Static Analysis Framework for Native CPython Bytecode}
\newcommand{\papertitleshort}{CPyGraph: A Version-Aware Static Analysis Framework for Native CPython Bytecode}
\newcommand{\tech}{\mbox{\textnormal{\textsc{CPyGraph}}}\xspace}

\newcommand{\bench}{\mbox{\textnormal{\textsc{PYGBench}}}\xspace}

\newcommand{\paperkeywords}{Python, CPython bytecode, static analysis, points-to analysis, call graph, control-flow graph, data-dependence graph}

\newcommand{\authorAname}{Baihong Chen}
\newcommand{\authorAaffil}{Utah State University}
\newcommand{\authorAemail}{b.chen@usu.edu}

\newcommand{\authorBname}{Wen Li}
\newcommand{\authorBaffil}{Utah State University}
\newcommand{\authorBemail}{awen.li@usu.edu}

\newcommand{\appendixcontent}{\appendix}

\newcommand{\artifact}[1]{%
  \href{https://github.com/awen-li/CPyGraph}{\underline{#1}}%
}

\setcopyright{acmcopyright}

\begin{document}

\title[\papertitleshort]{\papertitle}

\author{\authorAname}
\affiliation{%
  \institution{\authorAaffil}
  \city{Logan}
  \state{Utah}
  \country{USA}
}
\email{\authorAemail}
\authornote{First author}

\author{\authorBname}
\affiliation{%
  \institution{\authorBaffil}
  \city{Logan}
  \state{Utah}
  \country{USA}
}
\email{\authorBemail}
\authornote{Corresponding author}

\begin{CCSXML}
<ccs2012>
   <concept>
       <concept_id>10011007.10011074.10011099.10011692</concept_id>
       <concept_desc>Software and its engineering~Formal software verification</concept_desc>
       <concept_significance>500</concept_significance>
       </concept>
   <concept>
       <concept_id>10003752.10010124.10010131</concept_id>
       <concept_desc>Theory of computation~Program analysis</concept_desc>
       <concept_significance>500</concept_significance>
       </concept>
 </ccs2012>
\end{CCSXML}

\ccsdesc[500]{Theory of computation~Program analysis}
\ccsdesc[500]{Software and its engineering~Formal software verification}

\begin{abstract}
Static analysis of Python packages must recover both program structure and
object flow across first-class functions, dynamic dispatch, implicit protocol
calls, exceptions, closures, and module execution. Native CPython bytecode
provides the executable lowering of these behaviors, but its instruction,
call, stack, and exception representations change across releases. This
creates a need for a version-aware analysis foundation whose graph products
share the same bytecode identities and semantics.

We present \tech, a C++ framework for package-level analysis of native CPython
bytecode. Version-specific adapters expose stack, control, call, lexical,
protocol, and exception semantics through a shared interface while preserving
code-object identities and native bytecode offsets. An operand-stack-aware
Andersen points-to analysis and call graph grow together to a fixed point.
Their shared state supports exception-aware CFGs, block-level CDGs, and
interprocedural DDGs, with optional function-level flow, context, and bounded
path sensitivity. The framework also records unresolved dynamic behavior
through typed coverage summaries.

We evaluate \tech with \bench, 201 package-level programs and 1,733 fixed
candidates. On CPython 3.10, the default analysis reaches 91.40\% candidate
precision and 100\% recall; complete sensitivity reaches 94.97\% precision
with the same recall. Across CPython 3.10--3.14, 1,328 of 1,334
version-invariant queries agree, and \tech matches PyCG on its 112-program
call-graph benchmark.
\end{abstract}

\keywords{\paperkeywords}

\maketitle

\section{Introduction}\label{sec:intro}

Control-flow graphs (CFGs), control-dependence graphs (CDGs), points-to
information, call graphs (CGs), and data-dependence graphs (DDGs) are common
foundations for software engineering
tasks~\cite{kildall1973,andersen1994,grove1997callgraph,ferrante1987pdg}.
Constructing these relations consistently for Python is difficult.
Functions and classes are ordinary values, imports execute module bodies,
closures share lexical storage, and simple expressions can invoke
user-defined protocol methods. A useful framework must therefore recover both
program structure and the flow of objects through the operand stack, heap,
calls, and exceptional control flow~\cite{python-execution-model,
python-import-system,python-datamodel}.

Native CPython bytecode is an attractive analysis boundary. It is the
representation executed by CPython and records compiler-resolved control
transfers and operations as native instruction sequences~\cite{python-dis}.
It is also a practical artifact: Python package distributions include bytecode
without artifact-local source~\cite{chen2026beyondsource}. Source retains
names and intent; bytecode provides the compiler's executable lowering and
native execution locations.
CPython bytecode is version-specific. CPython treats bytecode as an
implementation detail and does not guarantee stability across
releases~\cite{python-dis}. Recent versions alter opcode numbers, call
sequences, exception tables, inline caches, and stack
effects~\cite{pep659}. A version-aware framework therefore needs a
release-specific decoding boundary while keeping stable native identities for
downstream analyses.

Python analysis systems provide source-level graph construction, points-to
analysis, abstract interpretation, and bytecode type analysis.
\texttt{codeanalyzer-python} emits call, control, and dependence
relations~\cite{canpy}; Scalpel provides reusable source analyses~\cite{li2022scalpel};
PoTo implements Andersen-style points-to analysis~\cite{rakamnouykit2025poto};
MOPSA provides flow- and context-sensitive abstract interpretation~\cite{monat2020python};
and Pytype analyzes CPython bytecode for type inference and checking~\cite{pytype}.
\tech focuses on a version-aware family of package-level analyses over native
code objects, with all products sharing one bytecode-semantic interface.

We address this need with \tech, a C++ library for package-level analysis
of native CPython code objects. Its adapters decode version-specific bytecode and expose the stack, control,
call, lexical, protocol, and exception semantics used by the shared analyses.
Each semantic record remains attached to its code object and native bytecode
offset. The CFG, CDG, PTA, CG, and DDG implementations use the same adapter
interface across releases.

The central analysis is an operand-stack-aware variant of inclusion-based Andersen
points-to analysis~\cite{andersen1994}. Points-to propagation and call-graph
construction proceed together. Resolving a callable activates the callee's
constraints and connects arguments, returns, receiver objects, and closure
storage. The resulting facts can resolve more callees, following the
established on-the-fly coupling of points-to and call-graph
construction~\cite{grove1997callgraph,lhotak2003spark}. The package-level API reuses this fixed point across the exposed products,
giving clients a consistent view of interprocedural behavior. Function-level policies optionally distinguish
flow locations, bounded call strings, and bounded acyclic path partitions
inside the points-to analysis~\cite{sharir1981,mauborgne2005trace}.

\tech records unresolved behavior explicitly. User-defined special methods
resolved in the package receive concrete call edges. Remaining dynamic
alternatives are represented by interned groups that retain identities such as
\texttt{\_\_exit\_\_} or \texttt{\_\_add\_\_}. Each analysis product also
carries a compact coverage summary that separates concrete facts from typed or
conservative unresolved behavior.
This article makes the following contributions:

\begin{itemize}[leftmargin=1.4em]
  \item A bytecode-native, operand-stack-aware Andersen points-to analysis for Python
        objects, fields, closures, modules, bound methods, and implicit protocol
        calls. Points-to propagation and call-graph construction share one
        fixed point, and clients can explicitly configure flow sensitivity,
        context sensitivity, and bounded path partitioning for selected functions.
  \item A version-adapter architecture that localizes raw opcode, stack, call,
        jump, lexical-access, protocol, and exception-table differences while
        preserving code-object identities and native bytecode offsets. The
        implementation ships adapters for CPython 3.10--3.14.
  \item An integrated C++ analysis framework that derives exception-aware CFGs,
        block-level CDGs, points-to sets, CGs, and interprocedural DDGs through
        stable APIs, together with \bench: 201 package cases and explicit,
        machine-readable ground truth for all five products.
\end{itemize}

Our semantic evaluation uses a fixed universe of 1,733 expected and
fault-revealing negative candidates. On CPython 3.10, the default insensitive
analysis reaches 91.40\% candidate precision and 100\% recall. Complete sensitivity
reports every one of 1,020 expected facts and rejects 659 of 713 negative
candidates, reaching 94.97\% candidate precision and 100\% recall. The control-dependence graph reaches 100\% candidate precision and recall, and
the call graph reaches 98.9\% candidate precision and 100\% recall. The
82.1\% points-to candidate precision accounts for most false positives. An
explicit configuration that assigns all three attributes to every function
matches complete sensitivity. Across CPython 3.10--3.14, insensitive candidate
precision stays within 0.35 percentage points and recall remains 100\%. Predictions agree for 1,328 of 1,334 version-invariant queries; the six
differences are CFG queries involving exception and cleanup lowering. On
PyCG's independent benchmark, \tech matches its 111 complete and 103 sound
call-graph cases. The real-package study completes 2,462 of 2,500 packages;
all 732 supported bundled \texttt{.pyc} files produce all five analysis products.

\noindent\textbf{Open science.}
The source, tests, benchmark programs, machine-readable ground-truth files, and
runner are available in the \artifact{CPyGraph repository}. The evaluation
includes semantic accuracy, an external call-graph comparison, cross-version
behavior, controlled ablations, real-package efficiency, and direct analysis
of bundled \texttt{.pyc} files (\cref{sec:eval}).

\section{Background and Motivation}\label{sec:background}

\subsection{CPython Code Objects and Stack Semantics}

CPython compiles a module and each nested function, lambda, comprehension,
class body, or generator into a code object. A code object contains bytecode,
constants, names, local and free-variable metadata, and exception metadata.
Instructions exchange values through an operand stack. A call target,
receiver, and arguments may therefore be separated by stack shuffles,
attribute loads, keyword markers, or version-specific call preparation.
The analysis therefore propagates abstract stack states along normal and
exceptional CFG edges~\cite{python-datamodel,python-dis}.
Bytecode is versioned. For example, CPython 3.10 represents calls and protected
regions differently from the 3.11+ instruction and exception-table design.
Later releases add fused local operations and change cache layout. The
official disassembler documents additions and removals between releases and
warns that raw adaptive cache data can resemble ordinary
instructions~\cite{python-dis,pep659}. These changes affect parsing, jump targets,
stack height, and therefore every downstream relation.

\subsection{Why a Bytecode-Level Foundation?}

\noindent\textbf{Executable representation.}
Compilation creates analysis identities and semantics such as code objects,
native bytecode offsets, operand-stack effects, exception regions, and inline
caches. These facts are the direct input to bytecode-level clients. Compilation
can also change program structure
across releases. For example, CPython 3.12 can inline a comprehension that an
earlier release represents as a separate code object~\cite{pep709}. \tech
therefore uses the selected CPython compiler and its native code objects as the
analysis boundary.

\vspace{3pt}
\noindent\textbf{Implicit Python behavior.}
Imports execute module bodies, closures share cells, and descriptors,
operators, iteration, context managers, and exception cleanup can invoke
functions not written as explicit source calls~\cite{python-datamodel}. The AST
identifies the language construct, while CPython bytecode records its lowering
as stack, call, and control-flow operations. Analyzing this representation lets
all products refer to the same executed operations and compiler-selected normal
and exceptional paths. Points-to reasoning resolves dynamic dispatch over this
common execution representation.

\vspace{3pt}
\noindent\textbf{Native bytecode artifacts.}
CPython also supports loading \texttt{.pyc} files without matching source
through its sourceless loader~\cite{python-importlib}. The header magic number
and marshaled code-object format make minor-version matching part of artifact
loading~\cite{pep552,python-marshal}. An empirical study of 1,034,843 PyPI
artifacts found 7,388 artifacts containing bytecode, including 28,193
\texttt{.pyc} files without matching source in the same
artifact~\cite{chen2026beyondsource}. These artifacts provide a direct use case
for an analysis path that accepts native bytecode.
These properties motivate a native-bytecode analysis boundary. It
accepts artifacts without matching source and preserves code-object identities,
native bytecode offsets, stack effects, and exception regions for downstream
clients. RQ1 evaluates the modeled language semantics, RQ2 evaluates
version-dependent lowering, RQ3 isolates key design choices, and RQ5 analyzes
bundled \texttt{.pyc} files directly.

\subsection{Motivating Example}\label{sec:motiexample}

\begin{figure}[t]
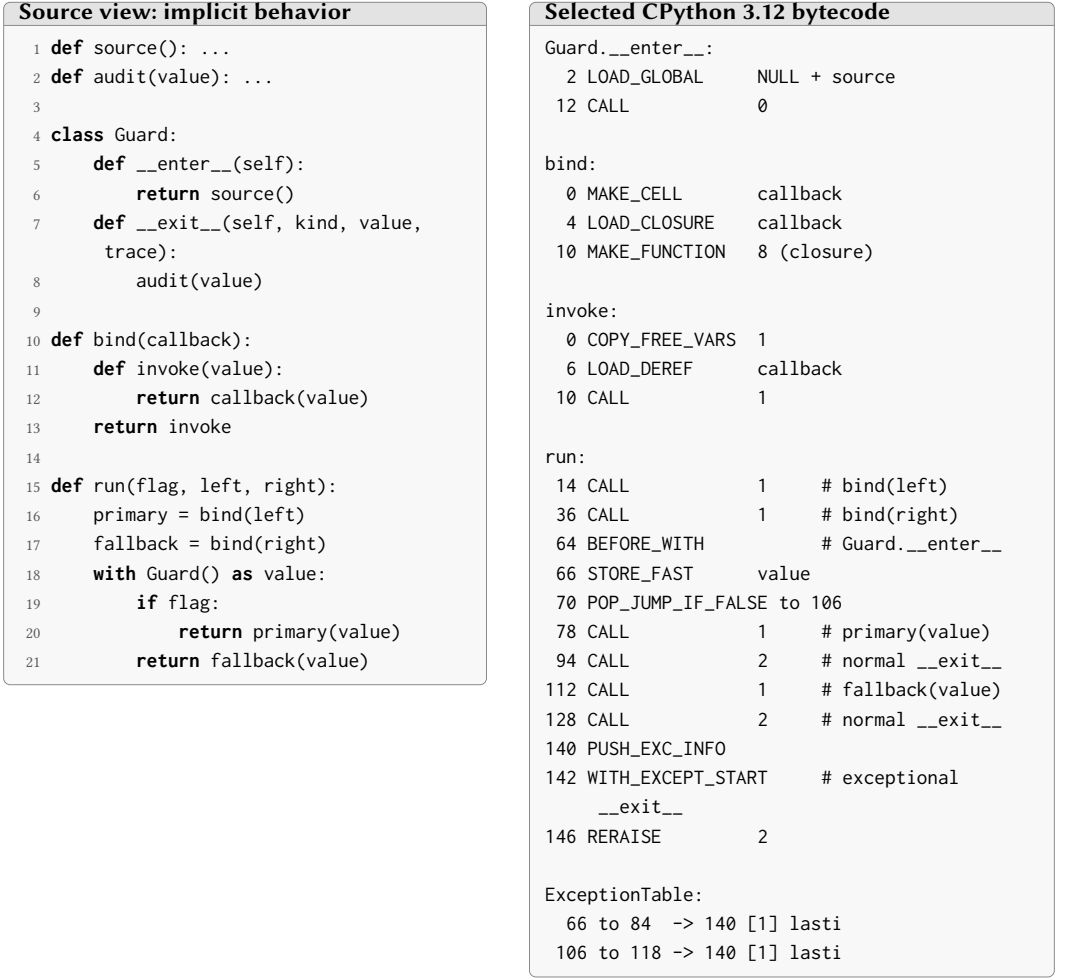

\centering
\begin{minipage}[t]{0.46\textwidth}
\vspace{0pt}
\begin{tcolorbox}[codepiece,title={\sffamily\bfseries Source view: implicit behavior},
  fonttitle=\sffamily\small,colbacktitle=black!11,coltitle=black]
\begin{lstlisting}[language=Python,numbers=left,numberstyle=\tiny\color{codecomment},
  numbersep=4pt,xleftmargin=1.5em,basicstyle=\ttfamily\footnotesize]
def source(): ...
def audit(value): ...

class Guard:
    def __enter__(self):
        return source()
    def __exit__(self, kind, value, trace):
        audit(value)

def bind(callback):
    def invoke(value):
        return callback(value)
    return invoke

def run(flag, left, right):
    primary = bind(left)
    fallback = bind(right)
    with Guard() as value:
        if flag:
            return primary(value)
        return fallback(value)
\end{lstlisting}
\end{tcolorbox}
\end{minipage}
\hfill
\begin{minipage}[t]{0.5\textwidth}
\vspace{0pt}
\begin{tcolorbox}[codepiece,title={\sffamily\bfseries Selected CPython 3.12 bytecode},
  fonttitle=\sffamily\small,colbacktitle=black!11,coltitle=black]
\begin{lstlisting}[basicstyle=\ttfamily\footnotesize]
Guard.__enter__:
  2 LOAD_GLOBAL     NULL + source
 12 CALL            0

bind:
  0 MAKE_CELL       callback
  4 LOAD_CLOSURE    callback
 10 MAKE_FUNCTION   8 (closure)

invoke:
  0 COPY_FREE_VARS  1
  6 LOAD_DEREF      callback
 10 CALL            1

run:
 14 CALL            1     # bind(left)
 36 CALL            1     # bind(right)
 64 BEFORE_WITH           # Guard.__enter__
 66 STORE_FAST      value
 70 POP_JUMP_IF_FALSE to 106
 78 CALL            1     # primary(value)
 94 CALL            2     # normal __exit__
112 CALL            1     # fallback(value)
128 CALL            2     # normal __exit__
140 PUSH_EXC_INFO
142 WITH_EXCEPT_START     # exceptional __exit__
146 RERAISE         2

ExceptionTable:
  66 to 84  -> 140 [1] lasti
 106 to 118 -> 140 [1] lasti
\end{lstlisting}
\end{tcolorbox}
\end{minipage}
\caption{The selected CPython 3.12 bytecode exposes protocol calls,
compiler-created control flow, closure storage, two callback contexts, the
implicit \texttt{\_\_enter\_\_} result, normal cleanup calls, and a shared
exceptional cleanup target. Other CPython versions use different native
instruction sequences.}
\Description{Two grayscale code panels show a Python program and selected
CPython 3.12 bytecode. The bytecode exposes closure storage, implicit
context-manager calls, branch-dependent callback invocation, and normal and
exceptional cleanup.}
\label{fig:motivating}
\end{figure}

\cref{fig:motivating} shows a small example that combines several Python
features relevant to bytecode analysis. The function \texttt{run} creates two
closures through \texttt{bind}, enters a \texttt{Guard} context, and invokes
one of the closures according to \texttt{flag}. The corresponding CPython
3.12 bytecode makes the closure-cell operations, the implicit
\texttt{\_\_enter\_\_}/\texttt{\_\_exit\_\_} calls, the branch-dependent
callback invocation, and the exceptional cleanup path explicit. The example
illustrates three requirements for a native-bytecode analysis foundation and
how \tech addresses them.

\vspace{3pt}
\noindent\textbf{Implicit executable behavior.}
The \texttt{with} statement invokes \texttt{\_\_enter\_\_} and
\texttt{\_\_exit\_\_} through Python's context-manager protocol. CPython
records these calls and their cleanup paths with \texttt{BEFORE\_WITH},
\texttt{WITH\_EXCEPT\_START}, and the exception table, including the
exceptional path taken when a callback raises.
The \emph{Version-Aware Bytecode Adapters} decode these native operations and
retain their code-object identities and native bytecode offsets.

\vspace{3pt}
\noindent\textbf{Context-dependent closures.}
Both calls to \texttt{bind} execute the same nested code object, but they
create different closure cells for \texttt{callback}. Merging the two cells
makes \texttt{primary} appear to call \texttt{right} and
\texttt{fallback} appear to call \texttt{left}. The relation between \texttt{invoke} and its parent is therefore needed to
recover both targets. The \emph{Shared Bytecode Analyses} represent closure cells explicitly and
resolve points-to facts and call edges together. Through the
\emph{Explicit Function-Level Sensitivity Configuration}, the analysis can
distinguish the two \texttt{bind} activations when a client needs these
precise targets.

\vspace{3pt}
\noindent\textbf{Cross-analysis semantic chains.}
The value returned by \texttt{source} crosses the implicit
\texttt{\_\_enter\_\_} call and becomes local \texttt{value}. The branch then
passes this value to either \texttt{primary} or \texttt{fallback}, while the
closure cell determines whether the final callback is \texttt{left} or
\texttt{right}. If a callback raises, control transfers through the
exceptional cleanup path and invokes \texttt{\_\_exit\_\_}, whose
\texttt{value} argument then reaches \texttt{audit}. These relations cross
analysis boundaries, so the CFG, CDG, PTA, CG, and DDG must use the same call
activations, points-to objects, native blocks, and unresolved groups. The
\emph{Public Analysis Products} reuse these facts to produce a CG with the
protocol and callback edges, a CFG with both normal and exceptional cleanup
paths, a CDG with the two branch outcomes, and a DDG that preserves the
corresponding data flows.
The example identifies the native operations that require modeling, the facts
shared across analyses, and the unresolved dynamic behavior that must remain
visible to clients.
These requirements lead directly to the version-aware adapters, shared
bytecode analyses, and public analysis products in
\cref{fig:architecture}.

\section{The {\tech} Framework}\label{sec:tech}
\subsection{Framework Overview}\label{ss:overview}

\tech has three design goals. First, it accepts native code objects directly.
Second, release-specific bytecode rules are localized in version adapters.
Third, all analysis products share the same object, call, and control-flow
semantics.
\cref{fig:architecture} shows the overall workflow. A client provides either
a source package or compatible native bytecode. Source packages are compiled
into native code objects, while bundled \texttt{.pyc} files or existing code
objects are loaded directly. A version-matched adapter decodes the native
instructions and exposes their stack, call, jump, exception, lexical-access,
and protocol semantics through a shared bytecode-semantic interface while
preserving code-object identities and native bytecode offsets.
The shared analyses operate on this interface. The CFG recovers normal and
exceptional control flow. The Andersen-style PTA tracks objects through the
operand stack, locals, fields, modules, closures, and call boundaries. PTA and
CG construction proceed together: newly resolved callees activate their
bodies, and the resulting constraints can expose additional call targets until
the two analyses reach a fixed point. An explicit function-level configuration
can refine selected PTA activations. The CFG then supports CDG construction,
while CFG facts, points-to objects, and resolved call activations are combined
to construct the interprocedural DDG.

\begingroup
\setlength{\intextsep}{10pt}
\begin{figure}[H]
\centering
\includegraphics[width=.65\linewidth]{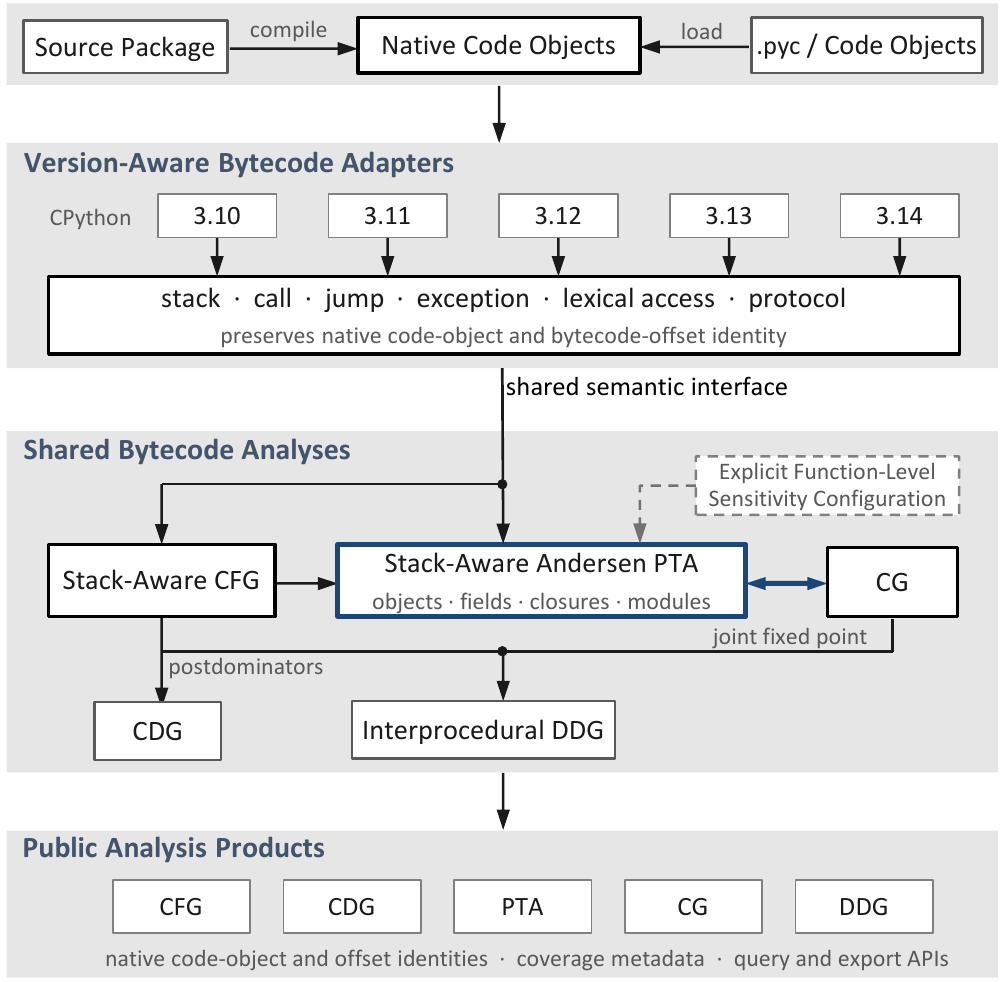}
\caption{Architecture and analysis workflow of \tech. Version-aware adapters
provide a shared bytecode-semantic interface used by the coordinated CFG,
PTA/CG, CDG, and DDG analyses.}
\label{fig:architecture}
\Description{A grayscale architecture diagram. Source packages or native
bytecode are converted to native code objects and processed by version-aware
adapters for CPython 3.10 through 3.14. The shared analyses construct CFG,
PTA, CG, CDG, and DDG products, with an explicit function-level sensitivity
configuration for PTA.}
\end{figure}
\endgroup

The resulting CFG, CDG, PTA, CG, and DDG share the same code-object
identities, native bytecode offsets, abstract objects, call activations, typed
unresolved groups, and coverage summaries. Clients query these products through
one package analysis state.

\subsection{Version-Aware Bytecode Adapters}
\label{sec:design:adapters}

The Version-Aware Bytecode Adapters in \cref{fig:architecture} combine
the package front end and version adapter. Together, they produce the semantic package $P_V$, a deterministic set of
semantic code objects for CPython version $V$. Each semantic instruction
retains its module, code-object identity, and native
bytecode offset. Downstream analyses consume these semantic records through the
shared bytecode-semantic interface.
Opcode values and cache layouts change between releases. Call and stack conventions may also
change when source syntax stays the same. Protected regions moved from setup
instructions to compact exception tables, while adaptive instructions added
inline cache entries~\cite{python-dis,pep659}.

The bundled package tools take source packages, discover modules in canonical
order, and compile them with the same minor version against which \tech is
built. The library also exposes the matching native-code-object loader, so a client
with compatible \texttt{.pyc} artifacts can load them directly~\cite{python-importlib,pep552,python-marshal}. In both routes, module bodies are
distinguished from nested code objects: modules are execution roots, whereas functions, lambdas,
comprehensions, and class bodies become reachable through Python semantics.

\Cref{alg:normalize-package} gives the boundary algorithm. A release adapter
owns the wordcode parser, opcode table, stack effects, branch and call layout,
lexical-access interpretation, protocol tags, and exception decoder. The
front end assigns dense numeric identities by walking each module root and its
nested code-object constants. A normalized exception tuple has the form
$(s,e,h,d,x,\ell)$. An instruction in $[s,e)$ may transfer to handler $h$.
The handler restores stack depth $d$, pushes $x$ exception values, and may
receive the last-instruction offset $\ell$.
The algorithm first selects the adapter and acquires
code objects for the same CPython minor version
(lines~\ref{alg:normalize:adapter}--\ref{alg:normalize:acquire}). It orders
module and nested code objects before assigning their stable identities
(lines~\ref{alg:normalize:order}--\ref{alg:normalize:identity}). Each raw
instruction is then lifted while retaining its native bytecode offset
(line~\ref{alg:normalize:lift}). The adapter decodes exception regions and
validates their instruction, jump, and stack references
(lines~\ref{alg:normalize:exceptions}--\ref{alg:normalize:validate}). The
result records every semantic code object and identifies the module roots
(lines~\ref{alg:normalize:package}--\ref{alg:normalize:return}).

\SetAlCapFnt{\scriptsize}
\SetAlCapNameFnt{\scriptsize}

\begin{center}
\begin{minipage}{0.6\linewidth}  
\begin{algorithm}[H]
\scriptsize
\setlength{\algomargin}{0.35em}
\DontPrintSemicolon
\LinesNumbered
\SetAlgoNlRelativeSize{-1}

\caption{Normalize Version-Matched Package Bytecode}
\label{alg:normalize-package}

\SetKwFunction{SelectAdapter}{selectAdapter}
\SetKwFunction{AcquireCode}{acquireCodeObjects}
\SetKwFunction{CodePreorder}{codePreorder}
\SetKwFunction{AssignIdentity}{assignIdentity}
\SetKwFunction{ParseInstructions}{parseInstructions}
\SetKwFunction{LiftInstruction}{liftInstruction}
\SetKwFunction{DecodeExceptions}{decodeExceptionRegions}
\SetKwFunction{ValidateCode}{validateCodeObject}
\SetKw{Return}{return}

\KwIn{$X$: package source or native code objects; $V$: CPython minor version}
\KwOut{$P_V$: semantic package; $R$: module roots}

$A \leftarrow \SelectAdapter(V)$\;
\nllabel{alg:normalize:adapter}

$M \leftarrow \AcquireCode(X,V)$\;
\nllabel{alg:normalize:acquire}

$C \leftarrow \CodePreorder(M)$; $P_V,R\leftarrow\emptyset$\;
\nllabel{alg:normalize:order}

\ForEach{$c \in C$}{
  $(id_c,parent_c,module_c)\leftarrow\AssignIdentity(c)$\;
  \nllabel{alg:normalize:identity}

  $I_c\leftarrow\emptyset$\;

  \ForEach{$r \in \ParseInstructions(A,c)$}{
    $i \leftarrow \LiftInstruction(A,r,c.\mathit{metadata})$\;

    $i.origin\leftarrow\langle id_c,r.offset\rangle$;
    $I_c\leftarrow I_c\cdot i$\;
    \nllabel{alg:normalize:lift}
  }

  $E_c\leftarrow\DecodeExceptions(A,c)$\;
  \nllabel{alg:normalize:exceptions}

  $\ValidateCode(I_c,E_c)$\;
  \nllabel{alg:normalize:validate}

  $P_V\leftarrow P_V\cup
    \{\langle id_c,parent_c,module_c,I_c,E_c\rangle\}$\;

  \If{$c$ is a module body}{
    $R\leftarrow R\cup\{id_c\}$\;
  }
  \nllabel{alg:normalize:package}
}

\Return{$P_V,R$}\;
\nllabel{alg:normalize:return}

\end{algorithm}
\end{minipage}
\end{center}

The semantic instruction record carries the information used by shared
analyses: stack inputs and outputs, branch-specific effects, jump targets,
call arguments, constants, lexical access, import metadata, and a fixed-size
Python-protocol mask. Version directories produce these records through
declarative opcode tables and small release-specific decoders. For CPython
3.10--3.14, release-specific decoding remains within the corresponding adapter
directories and their conformance tests.
This adapter boundary gives downstream analyses the compiler facts they need
through one shared interface. Native bytecode offsets and code-object nesting
provide stable identities for graph nodes and reports. Normalized stack and
control effects support graph construction. Protocol tags identify operations
that may invoke Python user code. The loader rejects code objects from a
different minor version before decoding them.

\noindent\textbf{Running example.}
Consider \texttt{run} in \cref{fig:motivating} under CPython 3.12. We write
stacks from bottom to top and use short presentation names for the numeric
identities stored by \tech. Let $o_G$ be the \texttt{Guard} instance,
$m_x$ its bound \texttt{\_\_exit\_\_} method, and $v_e$ the result of
\texttt{\_\_enter\_\_}. The adapter produces the following shared states:
\[
 [o_G]
 \xrightarrow{\mathsf{EnterContext}@64}
 [m_x,v_e]
 \xrightarrow{\mathsf{StoreLocal}(value)@66}
 [m_x].
\]
After loading \texttt{flag}, the branch at offset 70 consumes the condition
on both successors but retains $m_x$ below it. The encoded table entry printed
as \texttt{66 to 84 -> 140 [1] lasti} becomes
$\langle66,86,140,1,2,\mathsf{true}\rangle$. Thus, a may-raise instruction in
$[66,86)$ reaches handler 140 with the one-item protected prefix restored.
Downstream CFG construction consumes this normalized exception region directly.
The same source is represented by different native instruction sequences in
other releases. CPython 3.10 maps \texttt{SETUP\_WITH} to the same
\textsc{EnterContext} operation and reconstructs the protected interval from
setup instructions. CPython 3.14 exposes two \texttt{LOAD\_SPECIAL}
operations for \texttt{\_\_exit\_\_} and \texttt{\_\_enter\_\_}, followed by
\texttt{CALL}. Its adapter preserves these explicit loads while producing the
same protocol identities and stack facts used by the shared analyses.
\Cref{app:version-reference} gives the full cross-version trace.
\subsection{Shared Bytecode Analyses}
\label{sec:design:python-semantics}

The normalized instruction stream next receives Python object semantics.
Imports execute module bodies; functions and classes flow as values; attribute
loads can create bound methods; closures share mutable cells; and ordinary
syntax can dispatch to methods such as \texttt{\_\_call\_\_},
\texttt{\_\_enter\_\_}, or \texttt{\_\_exit\_\_}. \tech centralizes these rules in one package-level Python semantic coordinator
so the inclusion solver and graph builders consume the same semantic
facts~\cite{python-execution-model,python-import-system,python-datamodel}.
The coordinator represents functions, classes, modules, instances, bound
methods, closure cells, and selected literals as numeric abstract objects.
Named fields are interned numeric IDs. A closure object maps each free-variable
slot to a cell object, and the cell's contents occupy a distinguished field.
Thus nested code remains an independent function while reads, writes, sibling
closures, and multi-level captures share the correct storage
identity~\cite{python-execution-model}.

Imports connect stable module objects and exported fields. Attribute lookup
uses the receiver objects and visible class definitions; loading a function
through an instance creates a bound-method object that retains both receiver
and function. Protocol instructions carry a method-family bit mask from the
adapter. The coordinator applies Python's ordered fallback over package-visible
receiver types. A resolved package method becomes a concrete callable object. A remaining
dynamic alternative becomes an interned typed group that preserves its protocol
family.

\vspace{5pt}
\noindent\textbf{Operand-stack inclusion constraints.}
The points-to analysis operates on abstract locations tied to bytecode
execution. Let $s_{p,k}$ denote operand-stack position $k$ at program point
$p$, $l_x$ a local or global slot, $c_x$ a closure cell, and $h(o,f)$ field
$f$ of abstract object $o$. Each semantic instruction exposes its stack inputs
and outputs through the adapter, and the coordinator generates inclusion
constraints over these locations. Representative rules are:
\[
\begin{array}{rcll}
\mathsf{LoadLocal}(x)   &:& Pt(l_x) \subseteq Pt(s_{out}), \\
\mathsf{StoreLocal}(x)  &:& Pt(s_{in}) \subseteq Pt(l_x), \\
\mathsf{LoadField}(f)   &:& Pt(h(o,f)) \subseteq Pt(s_{out})
                              & \text{for } o\in Pt(s_{recv}), \\
\mathsf{StoreField}(f)  &:& Pt(s_{val}) \subseteq Pt(h(o,f))
                              & \text{for } o\in Pt(s_{recv}).
\end{array}
\]
Function and class creation allocate abstract callable or class objects at the
creating bytecode location; closure construction additionally connects the
function object to the referenced cell objects. When a call target $g$ is
resolved, the call boundary adds
\[
 Pt(a_i)\subseteq Pt(p_i),\qquad
 Pt(r_g)\subseteq Pt(s_{result}),
\]
for actual/formal pairs $(a_i,p_i)$ and callee return location $r_g$; bound
methods additionally connect the retained receiver to the callee's receiver
parameter. These rules make stack flow, heap fields, closures, and calls part
of one inclusion domain. The detailed Python feature rules are summarized in
\cref{app:python-feature-reference}.

\SetAlCapFnt{\scriptsize}
\SetAlCapNameFnt{\scriptsize}
\begin{center}
\begin{minipage}{0.8\linewidth}
\begin{algorithm}[H]
\scriptsize
\setlength{\algomargin}{0.35em}
\DontPrintSemicolon
\LinesNumbered
\SetAlgoNlRelativeSize{-1}
\caption{Compute the Shared PTA/CG Fixed Point}
\label{alg:python-coordination}

\SetKwFunction{SeedRequests}{seedSemanticRequests}
\SetKwFunction{PropagateFacts}{propagateInclusions}
\SetKwFunction{ReadyRequests}{readyRequests}
\SetKwFunction{ApplyRule}{applyPythonRule}
\SetKwFunction{NewCallees}{newPackageCallees}
\SetKwFunction{ActivateBody}{activateBody}
\SetKwFunction{ConnectBoundary}{connectCallBoundary}
\SetKwFunction{UnresolvedSites}{unresolvedSites}
\SetKwFunction{InternGroup}{internTypedGroup}
\SetKw{Return}{return}

\KwIn{semantic package $P_V$, module roots $R$, sensitivity policy $S$}
\KwOut{$Pt$: points-to facts; $A$: call activations; $G$: concrete call edges;
$U$: typed unresolved groups}

$Pt,A,G,U\leftarrow\emptyset$;
$Q\leftarrow\SeedRequests(P_V,R)$\;
\nllabel{alg:coord:seed}
$changed\leftarrow\mathsf{true}$\;
\While{$changed$}{
  $changed\leftarrow\PropagateFacts(Pt)$\;
  \nllabel{alg:coord:propagate}
  \ForEach{$\rho\in\ReadyRequests(Q,Pt)$}{
    $(\Delta Pt,\Delta Q)\leftarrow\ApplyRule(\rho,Pt)$\;
    \If{$\Delta Pt\nsubseteq Pt\lor\Delta Q\nsubseteq Q$}{
      $Pt\leftarrow Pt\cup\Delta Pt$;
      $Q\leftarrow Q\cup\Delta Q$;
      $changed\leftarrow\mathsf{true}$\;
      \nllabel{alg:coord:rules}
    }
  }
  \ForEach{$(q,o)\in\NewCallees(Q,Pt)$}{
    $(a,Q_a)\leftarrow\ActivateBody(o,q,S)$\;
    \nllabel{alg:coord:activate}
    $Pt\leftarrow\ConnectBoundary(Pt,q,a)$\;
    \nllabel{alg:coord:boundary}
    $A\leftarrow A\cup\{a\}$; $G\leftarrow G\cup\{q\rightarrow o\}$;
    $Q\leftarrow Q\cup Q_a$; $changed\leftarrow\mathsf{true}$\;
    \nllabel{alg:coord:edge}
  }
}
\ForEach{$q\in\UnresolvedSites(Q,Pt)$}{
  $u\leftarrow\InternGroup(q.feature,q.domain)$;
  $U\leftarrow U\cup\{u\}$; attach $u.id$ to $q$\;
  \nllabel{alg:coord:unresolved}
}
\Return{$Pt,A,G,U$}\;
\nllabel{alg:coord:return}
\end{algorithm}
\end{minipage}
\end{center}

Algorithm~\ref{alg:python-coordination} seeds semantic requests from module
roots (line~\ref{alg:coord:seed}) and alternates inclusion propagation with
Python-specific request handling
(lines~\ref{alg:coord:propagate}--\ref{alg:coord:rules}). A newly resolved
package callable activates a sensitivity-specific body instance and connects
its receiver, arguments, captures, and return value to the caller
(lines~\ref{alg:coord:activate}--\ref{alg:coord:boundary}). The same step adds
the call edge and any requests exposed by the callee
(line~\ref{alg:coord:edge}). After the fixed point, remaining nonmaterialized
behavior is attached to its site as a typed unresolved group
(line~\ref{alg:coord:unresolved}). CFG, CDG, and DDG clients reuse the returned
identities and call activations.

Points-to facts and call edges grow in the same loop. Let $F$ be the finite
set of points-to memberships, semantic requests, activations, and concrete call
edges available for one package under policy $S$. The solver state is the
powerset lattice $(2^F,\subseteq)$ and joins use set union. Each rule adds
facts by set union. Context sensitivity uses call strings of length at most two, and path
partitioning creates at most the configured number of acyclic variants per
function; excess alternatives are joined back into the function summary. The
joint process is therefore monotone over a finite domain and reaches a fixed
point~\cite{cousot1977abstract}. At the fixed point, every package-defined target reachable through the modeled
Python rules has activated its body and contributed its constraints. Behavior beyond the modeled package domain is recorded as a typed group for
graph clients. This construction instantiates inclusion-based points-to analysis over
the shared Python object domain~\cite{andersen1994}.

\vspace{3pt}
\noindent\textbf{Running example.}
At offsets 14 and 36 in \texttt{run}, the two calls resolve to
\texttt{bind}. Activating \texttt{bind} exposes the nested \texttt{invoke}
code object and the captured \texttt{callback} slot. Under the selective
context policy used below, the coordinator creates a cell $c_L$ for the
activation receiving \texttt{left} and a cell $c_R$ for the activation
receiving \texttt{right}. It records
\[
 Pt(c_L.\mathit{contents})=\{o_{left}\},\qquad
 Pt(c_R.\mathit{contents})=\{o_{right}\}.
\]
The returned closures pair \texttt{invoke} with $c_L$ and $c_R$, respectively.
In \texttt{invoke}, \texttt{LOAD\_DEREF} reads \texttt{callback}; that cell
content becomes the callee of the call at offset 10. The same fixed point adds
the corresponding callback edge as soon as each cell content becomes
available.
The context-manager path uses the same mechanism. The semantic operation at
offset 64 loads the package-defined \texttt{Guard.\_\_enter\_\_} target and
keeps \texttt{Guard.\_\_exit\_\_} on the abstract stack. The return of
\texttt{source} flows through \texttt{\_\_enter\_\_} into local
\texttt{value}; the retained exit target reaches the normal cleanup calls.
Any remaining descriptor or runtime-replacement alternative stays attached to
that site as a typed context-protocol group. The detailed feature rules appear
in \cref{app:python-feature-reference}.

\subsection{Explicit Function-Level Sensitivity Configuration}
\label{sec:design:sensitivity}

\tech supports three PTA sensitivity levels:
\emph{insensitive} (the default), \emph{selective} (an explicit
function-level configuration), and \emph{complete} (all sensitivity
attributes applied to every function). Selective mode lets a client refine
selected functions.

Each selective entry associates a code-object ID with a sensitivity bit
mask. Flow sensitivity distinguishes local state at program points.
Context sensitivity uses a two-call-site call string~\cite{sharir1981} and
makes allocation and receiver state activation-specific. This bound separates
common closure and callback activations while keeping the context domain
finite. The \emph{Path} attribute enables bounded acyclic path partitioning and implies
flow sensitivity. Cyclic choices and alternatives beyond the configured bound
merge conservatively~\cite{mauborgne2005trace}. The partition key is the
bounded branch history; no path-condition solver is used.
\Cref{alg:selective-activation} shows how these attributes determine the
constraint instances created for a function.

\SetAlCapFnt{\scriptsize}
\SetAlCapNameFnt{\scriptsize}
\begin{center}
\begin{minipage}{0.6\linewidth}
\begin{algorithm}[H]
\scriptsize
\setlength{\algomargin}{0.35em}
\DontPrintSemicolon
\LinesNumbered
\SetAlgoNlRelativeSize{-1}
\caption{Instantiate a Function under Selective Sensitivity}
\label{alg:selective-activation}

\SetKwFunction{SensitivityAttrs}{sensitivityAttributes}
\SetKwFunction{TwoSiteContext}{twoSiteContext}
\SetKwFunction{PathVariants}{boundedAcyclicVariants}
\SetKwFunction{MergedLocals}{mergedLocals}
\SetKwFunction{PointLocals}{programPointLocals}
\SetKwFunction{SharedObjects}{sharedObjects}
\SetKwFunction{ContextObjects}{contextObjects}
\SetKwFunction{InstantiateConstraints}{instantiateConstraints}
\SetKwFunction{MergeBoundary}{mergeCallBoundary}
\SetKw{Return}{return}

\KwIn{$f$: code object; $q$: incoming call; $S$: policy; $B$: path bound}
\KwOut{$\mathcal{I}_f$: constraint instances; $\beta_f$: shared call boundary}

$K\leftarrow\SensitivityAttrs(S,f)$\;
\If{$\mathsf{Path}\in K$}{
  $K\leftarrow K\cup\{\mathsf{Flow}\}$\;
}
\nllabel{alg:sensitivity:attributes}

$\kappa\leftarrow
  \begin{cases}
    \TwoSiteContext(f,q), & \mathsf{Context}\in K\\
    0, & \text{otherwise}
  \end{cases}$\;
\nllabel{alg:sensitivity:context}

$\Pi\leftarrow
  \begin{cases}
    \PathVariants(f,B), & \mathsf{Path}\in K\\
    \{\epsilon\}, & \text{otherwise}
  \end{cases}$\;
\nllabel{alg:sensitivity:paths}

$L\leftarrow
  \begin{cases}
    \PointLocals(f), & \mathsf{Flow}\in K\\
    \MergedLocals(f), & \text{otherwise}
  \end{cases}$\;
$O\leftarrow
  \begin{cases}
    \ContextObjects(f,\kappa), & \mathsf{Context}\in K\\
    \SharedObjects(f), & \text{otherwise}
  \end{cases}$\;
\nllabel{alg:sensitivity:locations}

$\mathcal{I}_f\leftarrow
  \{\InstantiateConstraints(f,L,O,\pi)\mid\pi\in\Pi\}$\;
\nllabel{alg:sensitivity:instantiate}

$\beta_f\leftarrow\MergeBoundary(\mathcal{I}_f,q)$\;
\nllabel{alg:sensitivity:boundary}
\Return{$\mathcal{I}_f,\beta_f$}\;
\nllabel{alg:sensitivity:return}
\end{algorithm}
\end{minipage}
\end{center}

Algorithm~\ref{alg:selective-activation} first obtains the sensitivity
attributes assigned to the function and makes the \emph{Path} attribute imply
flow sensitivity (line~\ref{alg:sensitivity:attributes}). Functions with the
\emph{Context} attribute receive a bounded call-string context
(line~\ref{alg:sensitivity:context}), while functions with the \emph{Path}
attribute receive bounded acyclic path variants
(line~\ref{alg:sensitivity:paths}). Flow
sensitivity determines whether locals are distinguished by program point, and
context sensitivity determines whether abstract objects are activation-specific
(line~\ref{alg:sensitivity:locations}). The resulting domains and path variants
instantiate the function constraints (line~\ref{alg:sensitivity:instantiate}),
and all instances connect to the caller through one shared argument/return
boundary (lines~\ref{alg:sensitivity:boundary}--\ref{alg:sensitivity:return}).

Sensitive and insensitive functions therefore participate in the same PTA/CG
fixed point. An unlisted function uses the insensitive abstraction. Bounded
contexts and path variants keep the state space finite. Complete mode applies
all three sensitivity attributes to every function. Assigning the same
attributes explicitly in selective mode yields the same analysis configuration
and serves as a conformance check for the policy mechanism.

\vspace{3pt}
\noindent\textbf{Running example.}
Without context sensitivity, the two activations of \texttt{bind} share one
cell abstraction, so both callback sites may target both $o_{left}$ and
$o_{right}$. A selective policy marks only \texttt{bind} and
\texttt{invoke} as context sensitive. The calls at offsets 14 and 36 then
create $\kappa_L$ and $\kappa_R$, which distinguish cells $c_L$ and $c_R$.
The resulting targets are
\[
 T(\mathit{invoke},\kappa_L,10)=\{o_{left}\},\qquad
 T(\mathit{invoke},\kappa_R,10)=\{o_{right}\}.
\]
All other functions remain insensitive and exchange facts through the same
call boundary. This example uses context sensitivity for \texttt{bind} and
\texttt{invoke}; the \emph{Flow} and \emph{Path} attributes can be configured
independently for other functions.
\Cref{app:sensitivity} gives a boxed configuration example and the exact rules
for identifiers, attribute combinations, bounds, and analysis modes.

\subsection{Public Analysis Products}
\label{sec:design:products}

This component materializes CFG, CDG, PTA, CG, and DDG views from one package
analysis state. The combined operation solves the PTA/CG fixed point once and
then builds the requested analysis products. CFG and CDG consume semantic
instructions and normalized exception regions. DDG additionally consumes PTA
objects and resolved call activations. The resulting products share code-object
IDs, call activations, fields, closure storage, and unresolved-behavior
metadata.

Each graph view has a small interface:

\begin{itemize}[leftmargin=1.5em,itemsep=2pt,topsep=3pt,parsep=0pt]
  \item \textbf{CFG.} It consumes semantic instructions and normalized
  exception regions. It exposes blocks, instruction mappings, and normal,
  branch, exception, and resume edges.

  \item \textbf{PTA and CG.} They consume Python semantic constraints and the
  selected activation policy. They expose points-to memberships, concrete call
  targets, contexts, and activations.

  \item \textbf{CDG.} It consumes the refined CFG. It exposes postdominators
  and block dependences labeled with normal, true, false, exception, or resume
  outcomes.

  \item \textbf{DDG.} It consumes CFG value flow, PTA object identities, and
  resolved call activations. It exposes local, global, heap, collection,
  closure, actual/formal, and return/result dependences.
\end{itemize}

\Cref{alg:materialize-products} describes the combined API. The CFG starts
with leader partitioning and typed successor construction. A definite-assignment
step removes disproved local-load exception edges. The CDG computes standard
postdominators over a synthetic exit. Nonterminating regions receive a
synthetic-exit successor. Dependence edges retain the branch outcome. The DDG
uses a forward reaching-definition worklist. These constructions follow the
standard postdominator-based definition of control dependence and monotone
data-flow iteration~\cite{ferrante1987pdg,kildall1973}. It then joins
actual/formal and return/result boundaries with the retained call activations. PTA-resolved base
objects restrict heap-field candidates. Same-block strong updates remove
overwritten definitions when the address identity agrees.

\SetAlCapFnt{\scriptsize}
\SetAlCapNameFnt{\scriptsize}
\begin{center}
\begin{minipage}{0.6\linewidth}
\begin{algorithm}[H]
\scriptsize
\setlength{\algomargin}{0.35em}
\DontPrintSemicolon
\LinesNumbered
\SetAlgoNlRelativeSize{-1}
\caption{Materialize Coordinated Analysis Products}
\label{alg:materialize-products}

\SetKwFunction{StartProfile}{startProfile}
\SetKwFunction{SemanticFixedPoint}{ptaCgFixedPoint}
\SetKwFunction{RefineCfg}{refineCFG}
\SetKwFunction{BuildCdg}{postdominatorCDG}
\SetKwFunction{BuildLocalDdg}{intraproceduralDDG}
\SetKwFunction{AppendDdg}{appendDDGs}
\SetKwFunction{StitchDdg}{stitchCallsAndHeap}
\SetKwFunction{CoverageSummary}{summarizeCoverage}
\SetKwFunction{FinishProfile}{finishProfile}
\SetKw{Return}{return}

\KwIn{$P_V$: semantic package; $R$: entry roots; $S$: sensitivity policy}
\KwOut{$Pt,G,\mathit{CFG},\mathit{CDG},\mathit{DDG}$; per-product coverage
$\mathcal{C}$; profile $T$}

$\tau\leftarrow\StartProfile()$\;
$(Pt,A,G,U)\leftarrow\SemanticFixedPoint(P_V,R,S)$\;
\nllabel{alg:products:fixedpoint}
\ForEach{code object $f\in P_V$}{
  $\mathit{CFG}_f\leftarrow\RefineCfg(f.\mathit{cfg},f.\mathit{parameters})$\;
  \nllabel{alg:products:cfg}
  $\mathit{CDG}_f\leftarrow\BuildCdg(\mathit{CFG}_f)$\;
  \nllabel{alg:products:cdg}
  $D_f\leftarrow\BuildLocalDdg(\mathit{CFG}_f)$\;
  \nllabel{alg:products:localddg}
}
$\mathit{DDG}\leftarrow\AppendDdg(\{D_f\})$\;
$\mathit{DDG}\leftarrow\StitchDdg(\mathit{DDG},A,Pt)$\;
\nllabel{alg:products:stitch}
$\mathcal{C}\leftarrow\CoverageSummary(P_V,U,Pt,G,\mathit{CFG},\mathit{CDG},\mathit{DDG})$\;
\nllabel{alg:products:coverage}
$T\leftarrow\FinishProfile(\tau)$\;
\Return{$Pt,G,\mathit{CFG},\mathit{CDG},\mathit{DDG},\mathcal{C},T$}\;
\nllabel{alg:products:return}
\end{algorithm}
\end{minipage}
\end{center}

Algorithm~\ref{alg:materialize-products} first computes the shared PTA/CG
fixed point, including points-to facts, call activations, concrete
call edges, and unresolved groups (line~\ref{alg:products:fixedpoint}). It then
refines the per-code-object CFG and uses that CFG for CDG and local DDG
construction (lines~\ref{alg:products:cfg}--\ref{alg:products:localddg}). Call
activations and points-to facts connect the local DDGs across calls and heap
locations (line~\ref{alg:products:stitch}). Finally, the component records
coverage for each product and returns the shared profile
(lines~\ref{alg:products:coverage}--\ref{alg:products:return}).

The PTA/CG fixed point uses the conservative CFG produced during bytecode
lifting. The refinement in line~\ref{alg:products:cfg} is a product-level
post-pass that removes local-load exception edges disproved by definite
assignment. CFG, CDG, and DDG use the refined control flow. PTA and CG keep the
may result from the PTA/CG fixed point. The post-pass is downstream of PTA/CG,
so the fixed-point computation remains monotone.

Each package-level single-product endpoint follows the same coordination
contract and retains only its requested view. The CFG and CDG builders consume
adapter output, while their package-level endpoints reuse the shared PTA/CG
state. The DDG additionally consumes call activations and PTA object identities
for interprocedural and heap stitching. When a client requests several
products, the combined API solves the shared fixed point once and reuses it.

Each product also carries a constant-size coverage summary. The coverage
domain is the three-element lattice
\[
\mathsf{ConcreteOnly}
\sqsubset
\mathsf{TypedUnresolved}
\sqsubset
\mathsf{ConservativeTop},
\]
and joins select the least upper bound. A product is
\textsc{ConcreteOnly} when its materialized relation is complete within the
modeled package domain. \textsc{TypedUnresolved} means that remaining behavior
is limited to recorded feature or protocol groups, such as context cleanup or
binary dispatch. \textsc{ConservativeTop} means that any compatible fact in
the client's finite candidate domain may still hold. The package summary is
the join of its site-level coverage states.

These shared identities and coverage summaries keep the products consistent.
All views observe the same reachable bodies, implicit calls, objects, and call
activations. Stable IDs support direct cross-product queries, while coverage
summaries preserve unresolved behavior without expanding it into dense graph
edges. Nodes, edges, fields, contexts, and unresolved groups remain numeric;
names are decoded only by reports and DOT rendering.

\vspace{3pt}
\noindent\textbf{Running example.}
The fixed point above gives every product the same numeric identities. Let
$B_t$ contain the test at offset 70, $B_p$ and $B_f$ contain the two callback
calls, and $B_x$ contain the exceptional cleanup at offset 140. Representative
facts are:
\begin{itemize}[leftmargin=1.5em,itemsep=1pt,topsep=2pt,parsep=0pt]
  \item \textbf{PTA:} $Pt(c_L.\mathit{contents})=\{o_{left}\}$ and
        $Pt(c_R.\mathit{contents})=\{o_{right}\}$.
  \item \textbf{CG:} the calls at offsets 14 and 36 target \texttt{bind};
        $(\mathit{invoke},\kappa_L,10)$ targets \texttt{left}, while
        $(\mathit{invoke},\kappa_R,10)$ targets \texttt{right}. The retained
        cleanup callable targets \texttt{Guard.\_\_exit\_\_}.
  \item \textbf{CFG:} $B_t\xrightarrow{\mathsf{true}}B_p$ and
        $B_t\xrightarrow{\mathsf{false}}B_f$; may-raise operations in either
        arm have an exception successor to $B_x$.
  \item \textbf{CDG:} $B_p$ and $B_f$ depend on the true and false outcomes
        of $B_t$, respectively.
  \item \textbf{DDG:} the result of \texttt{source} reaches the return of
        \texttt{\_\_enter\_\_}, local \texttt{value}, and the selected
        callback argument. Each callback callee also depends on its distinct
        captured-cell content.
\end{itemize}

All five products therefore describe the same execution using the same
identities. Concrete package targets appear as graph edges, while unresolved
dynamic alternatives remain recorded as typed coverage groups.

\section{Implementation and Limitations} \label{sec:impl}

\subsection{Library and Package Pipeline} \label{sec:impl:components}

We implemented \tech in C++17. CMake builds static and shared libraries, and
clients can include the complete API or component headers for adapters, CFG,
PTA, CG, DDG, visualization, Python protocols, package analysis, and profiling.
The package pipeline has three explicit stages: compile, load, and analyze.
The compiler recursively discovers Python modules, supports conventional
\texttt{src/} layouts, and uses the interpreter selected when \tech is built.
The compiler rejects a selected interpreter whose major or minor version does
not match the CPython library linked into \tech. The loader preserves modules
and nested code objects, checks the \texttt{.pyc} magic header against that
linked runtime, and rejects truncated, incompatible, or non-code-object input
before lifting~\cite{python-importlib,pep552,python-marshal}.

Adapters for CPython 3.10, 3.11, 3.12, 3.13, and 3.14 live in separate
directories. Each directory contains a generated macro header for native opcode values and a
declarative opcode-to-semantics table. CMake discovers adapter directories and
generates the factory registry. The five supported releases use the same public
API and shared analyses. The implementation ships all five adapters, and
\cref{sec:rq2} evaluates semantic accuracy and consistency across them.
The graph representation uses numeric IDs for nodes, edges, contexts, fields,
code objects, and unresolved groups. Protocol alternatives are fixed-size bit
masks. Debug names live in intern tables and appear only in reports and DOT
output. The library exposes immutable graph views. DOT generators render all
node kinds as rectangles and decode numeric metadata at the presentation
boundary.

Profiling is enabled by default. Scoped regions record wall time, process CPU
time, and starting, ending, peak, and delta process RSS. A background thread
samples RSS every 5\,ms while at least one region is active. Clients can turn
profiling off through the same library API or a command-line flag. Optional
tools---\texttt{pygCG}, \texttt{pygCFG}, \texttt{pygCDG}, and
\texttt{pygDDG}---accept a
package, run the complete PTA/CG prerequisite, and emit the requested result
and profile as JSON.

\subsection{Scope and Limitations} \label{sec:impl:limits}

\tech is a may-analysis for package-defined Python code. Dynamic code
generation, unrestricted reflection, imports whose names are computed at run
time, and external or native library behavior can remain unresolved. Package module bodies define the analysis roots; imported modules
outside the selected package remain at this boundary. Coverage summaries record
these boundaries. A client projects typed or conservative unresolved behavior
onto the finite domain used by its analysis.

The framework models coroutine and suspension bytecode structurally. Async
scheduling and interleavings require a separate event and scheduler model
because event loops can run other tasks and callbacks whenever a task
awaits~\cite{python-asyncio}. The DDG contains data-dependence edges; control
dependence is provided separately by the CDG. The \emph{Path} attribute
partitions bounded acyclic choices inside PTA, and loops or excess variants
merge to guarantee termination.

Finally, CPython bytecode is implementation-specific. The implementation
supports CPython 3.10--3.14; other Python virtual machines require separate
semantics. A client analyzing a mixed-version corpus uses the \tech build that
matches each bytecode minor version. The input contract accepts bytecode that
the matching CPython runtime accepts. Malformed-but-parseable bytecode is not
part of the evaluation.

\section{Evaluation} \label{sec:eval}

We organize the evaluation around five research questions.
\begin{description}[style=unboxed,leftmargin=0cm]
\denseitems
\item[\textbf{RQ1:}] \emph{How accurately does \tech recover CFG, CDG, PTA, CG, and DDG facts?}
\item[\textbf{RQ2:}] \emph{How accurately and consistently does \tech analyze
programs across CPython 3.10--3.14?}
\item[\textbf{RQ3:}] \emph{How much do field modeling, exception handling,
PTA sensitivity, and interprocedural DDG contribute to the results?}
\item[\textbf{RQ4:}] \emph{What runtime and memory costs does \tech incur on
real Python packages?}
\item[\textbf{RQ5:}] \emph{Can \tech analyze bundled \texttt{.pyc} files directly?}
\end{description}

\subsection{Experimental Setup}\label{sec:eval-setup}

\subsubsection{Environment}

We use a 24-core Intel Core i9-13900F machine with 62\,GiB of memory, Ubuntu
22.04.5, and Linux 6.8.0. We build \tech with GCC 9.5.0 and CMake 3.29.6.
Separate CPython 3.10--3.14 environments provide a matching \tech build for
each version. RQ1 uses CPython 3.10; RQ2 uses all five versions. Each
package--configuration pair runs once, with a 30-minute timeout and a 32-GiB
memory limit. The semantic analysis is deterministic for a fixed package,
interpreter, and configuration. The same run produces all five analysis products
and records wall time, CPU time, and process-tree peak RSS. Peak RSS includes
compilation subprocesses.

\subsubsection{Benchmarks}

\noindent\textbf{\bench.}
\bench contains 201 small Python packages. Each case isolates a language
feature or analysis challenge and supplies ground truth for PTA, CG, CFG, CDG,
and DDG. The cases cover modules and imports; functions and closures; classes,
fields, descriptors, and metaclasses; implicit protocol calls; branches,
loops, exceptions, and cleanup; containers; and interprocedural flow. Async scheduling is not part of the benchmark. This feature-isolating design follows
prior Python call-graph, type-analysis, and points-to
microbenchmarks~\cite{salis2021pycg,venkatesh2024typeevalpy,prakash2019pointeval}.

The benchmark declares a fixed universe of 1,733 candidates: 1,020 expected
facts and 713 fault-revealing negative facts, with per-component counts in
\cref{tab:universe}. PTA candidates are points-to memberships. CG, CFG, CDG,
and DDG candidates are bounded semantic paths, including single-edge paths. A
path is present only when all of its edges are present. Expected and negative
candidates are authored with each case and frozen before the scored runs.
Negative candidates encode plausible alternatives for the same feature: a
wrong points-to object, callable, branch or handler, control relation, or data
producer. The candidate universe is finite and does not enumerate every absent
graph edge. Every case ships machine-readable ground truth, expected DOT
graphs, and PTA sets.

\vspace{3pt}
\noindent\textbf{PyCG call-graph benchmark.}
For an independent CG comparison, we use PyCG's 112 programs and published
ground truth~\cite{salis2021pycg}. We pin artifact commit
\texttt{3b37b54} and run both tools with CPython 3.10. This experiment compares
the call-graph edge relation shared by the two tools.

\begin{table}[H]
\centering
\caption{Fixed \bench candidate universe. Negative candidates are facts that
must not be reported; their presence counts as a false positive.}
\label{tab:universe}
\resizebox{0.5\linewidth}{!}{%
\begin{tabular}{lrrr}
\toprule
Component & Expected & Negative & Total \\
\midrule
PTA memberships & 133 & 140 & 273 \\
CG paths        & 182 & 230 & 412 \\
CFG paths       & 286 &  62 & 348 \\
CDG paths       & 153 & 218 & 371 \\
DDG paths       & 266 &  63 & 329 \\
\midrule
All graph paths & 887 & 573 & 1,460 \\
All candidates  & 1,020 & 713 & 1,733 \\
\bottomrule
\end{tabular}%
}
\end{table}

\noindent\textbf{Real-package benchmark.}
RQ4 uses 2,500 unique PyPI packages. We begin with the latest stable source
distributions released between September 6, 2023 and September 6, 2026 in
PyPI's public metadata~\cite{pypi-bigquery}. Eligible packages contain Python
source, support the assigned interpreter through
\texttt{Requires-Python}~\cite{python-core-metadata}, link to a recently active public repository, and
compile under that interpreter. For each CPython version, we divide eligible
packages into ten equal-count source-size strata and sample 50 packages per
stratum with seed 20260910. This yields 500 packages per version and 2,500 in
total, with unique package names and repositories. The frozen manifest pins
each release and archive hash. The corpus contains 162,962 Python files and
32.24 million source lines. RQ5 uses its fixed subset containing bundled
\texttt{.pyc} files.

\subsubsection{Metrics}

\noindent\textbf{Accuracy.}
Expected facts contribute true positives (TP) or false negatives (FN), while
negative facts contribute true negatives (TN) or false positives (FP). We sum
the counts over \bench before computing candidate precision
$TP/(TP+FP)$ and recall $TP/(TP+FN)$. Candidate precision is defined only over
the declared universe; facts outside that universe do not enter the score.
Unresolved summaries are projected onto compatible candidates, so conservative
fallbacks affect candidate precision. External runtime targets are outside the
package-level ground truth.
The PyCG benchmark uses its published case-level measures: a complete case has
no false-positive edge, and a sound case has no false-negative edge.
RQ2 reports accuracy per version and prediction consistency: the fraction of
version-invariant queries with identical predictions across all five versions.
RQ3 reports changes in TP, FP, FN, candidate precision, recall, runtime, and peak RSS
relative to the full system.

\vspace{3pt}
\noindent\textbf{Efficiency and applicability.}
RQ4 reports wall and CPU time and peak RSS. We summarize
package-level results with medians, 95th percentiles, and ranges, and use phase
times and graph sizes to explain cost. RQ5 reports bundled bytecode files,
recovered code objects, component completion, time, memory, and failure types.

\subsection{RQ1: Accuracy}\label{sec:rq1}

RQ1 reports both the default and the most precise built-in configurations. On
the CPython 3.10 candidate universe, the default insensitive analysis reaches
91.40\% candidate precision and 100\% recall (\cref{tab:sensitivity}). Complete
sensitivity reaches 94.97\% candidate precision and 100\% recall and provides the
component-level results below. RQ3 isolates the effect of the sensitivity
configuration.
\tech reports all 1,020 expected facts, so every component reaches 100\%
recall (\cref{tab:correctness}). This includes expected exceptional CFG paths,
typed control dependencies, implicit protocol calls, cross-module flow, closure
storage, and interprocedural dependences. These measurements characterize accuracy within the fixed
\bench candidate domain.
Candidate precision differs by component. CDG rejects all 218 negative paths.
CG has two false-positive paths and reaches 98.91\%. DDG reaches 97.79\%, CFG
reaches 94.39\%, and PTA reaches 82.10\%. PTA accounts for 29 of the 54 false
positives. This pattern matches the
pipeline design: points-to sets preserve alternatives at weakly updated
globals, fields, containers, and typed unresolved groups even when downstream
receiver, control-flow, or path-partition facts can reject incompatible graph
paths.

\begin{table}[H]
\centering
\caption{Complete-sensitivity accuracy on the fixed \bench candidate universe for all 201 cases.}
\label{tab:correctness}
\resizebox{0.5\linewidth}{!}{%
\begin{tabular}{lrrrrrr}
\toprule
Component & TP & TN & FP & FN & Cand. prec. & Recall \\
\midrule
PTA & 133 & 111 & 29 & 0 & 82.10\% & 100\% \\
CG  & 182 & 228 &  2 & 0 & 98.91\% & 100\% \\
CFG & 286 &  45 & 17 & 0 & 94.39\% & 100\% \\
CDG & 153 & 218 &  0 & 0 & 100\%   & 100\% \\
DDG & 266 &  57 &  6 & 0 & 97.79\% & 100\% \\
\midrule
All & 1,020 & 659 & 54 & 0 & 94.97\% & 100\% \\
\bottomrule
\end{tabular}
}%
\end{table}

\noindent\textbf{External call-graph comparison.}
We run \tech on PyCG's pinned 112-program micro-benchmark using its published
call-graph ground truth and scoring criteria. The comparison uses the
qualified-name call-edge relation shared by both analyses. As
\cref{tab:cg-comparison} shows, PyCG and both \tech configurations reach 111
complete cases and 103 sound cases. The \tech runner records every reported,
missing, and extra edge for each case.

\begin{table}[H]
\centering
\caption{Call-graph results on the independent PyCG micro-benchmark. A
complete case has no extra edge; a sound case has no missing edge.}
\label{tab:cg-comparison}
\resizebox{0.38\linewidth}{!}{%
\begin{tabular}{lrr}
\toprule
Tool/configuration & Complete & Sound \\
\midrule
PyCG & 111/112 & 103/112 \\
\tech insensitive & 111/112 & 103/112 \\
\tech complete & 111/112 & 103/112 \\
\bottomrule
\end{tabular}
}%
\end{table}

\noindent\textbf{Answer to RQ1.}
Across \bench's fixed 1,733-candidate universe, the default configuration
reaches 91.40\% precision and 100\% recall; complete sensitivity
reaches 94.97\% precision and 100\% recall. CG, CFG, CDG, and DDG
each exceed 94\% precision under complete sensitivity. PTA accounts
for the largest share of false positives. On the independent PyCG call-graph
benchmark, \tech matches PyCG's 111
complete and 103 sound cases out of 112.

\subsection{RQ2: Version Awareness}\label{sec:rq2}

The cross-version study compiles all 201 \bench cases with CPython 3.10, 3.11,
3.12, 3.13, and 3.14, then selects the matching \tech build and runs the shared
analysis. We compare semantic candidates across the version-matched oracles because the
same source can compile to different code-object and control-flow structures.
Each CPython version therefore has a version-matched oracle. This oracle keeps
the source-level intent fixed while expressing each fact with the semantic
identities produced by that version. The small changes in expected and negative
counts in \cref{tab:cross-version} come from candidates whose semantic identity
or label changes across releases; comprehension inlining from CPython 3.12 is
one such change~\cite{pep709}.

\begin{table}[H]
\centering
\caption{Cross-version insensitive-analysis results on \bench.}
\label{tab:cross-version}
\resizebox{0.5\linewidth}{!}{%
\begin{tabular}{lrrrr}
\toprule
CPython & Expected & Negative & Cand. prec. & Recall \\
\midrule
3.10 & 1,020 & 713 & 91.40\% & 100\% \\
3.11 & 1,021 & 712 & 91.08\% & 100\% \\
3.12 & 1,017 & 716 & 91.05\% & 100\% \\
3.13 & 1,017 & 716 & 91.05\% & 100\% \\
3.14 & 1,017 & 716 & 91.13\% & 100\% \\
\bottomrule
\end{tabular}
}%
\end{table}

\Cref{tab:cross-version} shows 100\% recall in every release, with candidate precision
within 0.35 percentage points across the five builds. The version matrix also runs the same sensitivity configurations used in RQ3.
Every configuration retains 100\% recall, and Explicit--All exactly matches
Complete in every version.
We also compare exact predictions across versions. A query is
version-invariant when the same canonical semantic identity has the same
expected/negative label in all five version-matched oracles. This rule selects
1,334 queries. The remaining 399 candidates change identity, label, or
availability in at least one release and are excluded from the consistency
denominator. Insensitive analysis gives the same prediction for 1,328 of the
1,334 invariant queries (99.55\%). The six mismatches are CFG queries involving exception or cleanup lowering.
CDG, PTA, CG, and DDG predictions are identical on their invariant queries in
this policy.

The experiment driver also measures extension surface directly from the
version directories. CPython 3.10 through 3.14 require 396, 158, 165, 178, and
230 nonblank, noncomment handwritten lines, respectively; generated opcode
headers contain 127, 110, 145, 154, and 162 entries. Relative to the preceding
release, the opcode-name sets add/remove 29/46, 53/18, 12/3, and 19/11 entries.
The artifact additionally records semantic-table entries, feature assignments,
and version-specific virtual overrides, keeping generated tables separate from
handwritten adaptation effort.

\noindent\textbf{Answer to RQ2.}
Across CPython 3.10--3.14, insensitive analysis retains 100\% recall and keeps
candidate precision within 0.35 percentage points. Exact predictions agree on
99.55\% of version-invariant queries, with six CFG mismatches. Each adapter added after the 3.10 foundation contains 158--230 handwritten
lines.

\subsection{RQ3: Design Contributions}\label{sec:rq3}

We use controlled ablations to isolate implemented design choices. The
experiments cover field modeling, exception-flow handling, PTA sensitivity,
and interprocedural DDG propagation. Version adaptation is evaluated directly
in RQ2.

\subsubsection{PTA sensitivity configuration}

\begin{table}[H]
\centering
\caption{Whole-dataset candidate precision under PTA sensitivity policies. All policies
have TP$=1{,}020$, FN$=0$, and therefore 100\% recall.}
\label{tab:sensitivity}
\resizebox{0.4\linewidth}{!}{%
\begin{tabular}{lrrr}
\toprule
Policy & TN & FP & Cand. prec. \\
\midrule
Insensitive        & 617 & 96 & 91.40\% \\
Uniform--Flow    & 622 & 91 & 91.81\% \\
Uniform--Context & 633 & 80 & 92.73\% \\
Uniform--Path    & 629 & 84 & 92.39\% \\
Explicit--All     & 659 & 54 & 94.97\% \\
Complete           & 659 & 54 & 94.97\% \\
\bottomrule
\end{tabular}
}%
\end{table}

Sensitivity changes the PTA abstraction, and the whole pipeline consumes its results.
\cref{tab:sensitivity} therefore reports the same 1,733-candidate universe for
every policy. Insensitive analysis keeps all expected facts and reaches 91.40\%
candidate precision. Applying flow sensitivity, context sensitivity, or bounded
path partitioning uniformly to every function removes
5, 16, and 12 false positives, respectively. Context sensitivity yields the largest single-attribute gain. The three
single-attribute reductions sum to 33 false positives, while applying all three
removes 42. The additional nine come from interactions: some call and data-flow
candidates disappear when multiple sources of merging are removed together.
Explicitly assigning all attributes to every function reaches the same result
as complete sensitivity.

Explicit--All and Complete assign the same attributes to every function through
two configuration paths. Their equal results confirm configuration equivalence.
Uniform--Flow, Uniform--Context, and Uniform--Path isolate one attribute at a
time across all functions. This experiment evaluates uniform policies; sparse
policies use the same explicit function-level configuration described in
\cref{sec:design:sensitivity}.

\subsubsection{Controlled ablation protocol}
For each remaining factor, a compile-time experiment option changes one
implemented mechanism at a time: field names collapse into the summary field, exception regions are omitted during
CFG construction, or actual/formal and return/call DDG stitching is disabled.
The public ablation driver builds these three variants plus the unmodified
system and reruns the complete \bench candidate universe under insensitive
PTA. \Cref{tab:ablations} reports the direct change in the primary component and
the measured propagation to the complete candidate universe. Runtime and peak
RSS vary by less than 0.5 seconds and 0.8 MiB in these small cases, so these
controls are used to isolate accuracy effects.

\begin{table}[H]
\centering
\caption{Controlled ablations on \bench under insensitive PTA. Primary
$\Delta$FP/$\Delta$FN are measured on the named component; the last column
shows the observed whole-suite propagation.}
\label{tab:ablations}
\resizebox{0.78\linewidth}{!}{%
\begin{tabular}{llrrrrl}
\toprule
Removed mechanism & Component & Full prec. & Ablated prec. & $\Delta$FP & $\Delta$FN & Whole-suite effect \\
\midrule
Field model & PTA & 73.08\% & 71.89\% & +3 & 0 & FP: 96$\rightarrow$122 \\
Exception flow & CFG & 94.39\% & 89.74\% & +15 & +6 & FN: +12 \\
Interprocedural flow & DDG & 97.44\% & 97.21\% & 0 & +22 & FN: +22 \\
\bottomrule
\end{tabular}
}%
\end{table}

Collapsing fields creates three additional PTA false positives. Its effects
propagate through the coupled pipeline, increasing overall false positives
from 96 to 122. Removing exception regions loses six CFG facts and creates 15
additional CFG false positives; downstream products raise the overall loss to
12 expected facts. Removing interprocedural DDG stitching loses 22 DDG facts, which are also the 22 whole-suite false negatives. The ablation set covers the three mechanisms reported in \cref{tab:ablations}.

\noindent\textbf{Answer to RQ3.}
Uniform flow sensitivity, context sensitivity, and bounded path partitioning
each improve whole-suite candidate precision
while retaining every expected fact; explicitly applying all three attributes
to every function removes 42 of 96 false positives. The field, exception-flow,
and interprocedural-DDG controls show measurable losses in their primary
components.

\subsection{RQ4: Efficiency}\label{sec:rq4}

We run the complete pipeline once on each of the 2,500 packages. 
End-to-end measurements include
package compilation and all child processes. Component measurements come from
the in-process profile of the same run.

\noindent\textbf{End-to-end cost.}
\Cref{tab:rq4-end-to-end} reports the full cost distribution. A package takes
0.95 seconds and 28.40\,MiB at the median. At the 95th percentile, it takes
45.97 seconds and 231.95\,MiB. The maximum observed cost is 1,782.28 seconds
and 1,457.03\,MiB.

\begin{table}[H]
\centering
\caption{End-to-end cost over successful real-package analyses.}
\label{tab:rq4-end-to-end}
\resizebox{0.45\linewidth}{!}{%
\begin{tabular}{lrrr}
\toprule
Metric & Median & P95 & Maximum \\
\midrule
Wall time (s) & 0.95 & 45.97 & 1,782.28 \\
CPU time (s) & 0.94 & 46.12 & 1,801.93 \\
Peak RSS (MiB) & 28.40 & 231.95 & 1,457.03 \\
\bottomrule
\end{tabular}%
}
\end{table}

\begin{figure}[H]
\centering
\includegraphics[width=\linewidth]{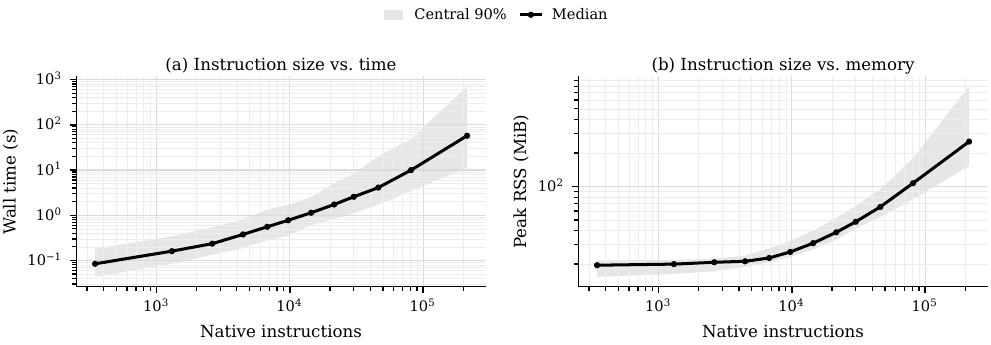}
\caption{End-to-end cost versus instruction count. Curves show medians and
central 90\% ranges over successful analyses.}
\Description{Two grayscale plots show binned median wall time and peak RSS
with central 90 percent ranges as native instruction count increases.}
\label{fig:rq4-end-to-end-scaling}
\end{figure}

\noindent\textbf{End-to-end scaling.}
We order packages by the number of native bytecode instructions.
\Cref{fig:rq4-end-to-end-scaling} summarizes 12 equal-count bins using medians
and central 90\% ranges; \cref{tab:rq4-size-scaling} gives a compact
five-group summary. Both show monotonic growth in the medians. From the
smallest to the largest group, median time increases from 0.13 to 15.99
seconds and median peak RSS increases from 19.82 to 132.75\,MiB. Across
individual packages, the Spearman correlations between instruction count and
wall time and peak RSS are 0.959 and 0.991, respectively.

\begin{table}[H]
\centering
\caption{End-to-end scaling by native-instruction quintile.}
\label{tab:rq4-size-scaling}
\resizebox{0.7\linewidth}{!}{%
\begin{tabular}{lrrr}
\toprule
Quintile (instruction range) & Median instructions & Median time (s) & Median RSS (MiB) \\
\midrule
Q1 (4--2,511) & 1,045 & 0.13 & 19.82 \\
Q2 (2,512--7,480) & 4,681 & 0.40 & 21.49 \\
Q3 (7,483--19,144) & 11,707 & 0.95 & 28.25 \\
Q4 (19,183--47,722) & 29,272 & 2.50 & 48.11 \\
Q5 (47,742--1,242,497) & 107,045 & 15.99 & 132.75 \\
\bottomrule
\end{tabular}%
}
\end{table}

\begin{figure}[H]
\centering
\includegraphics[width=\linewidth]{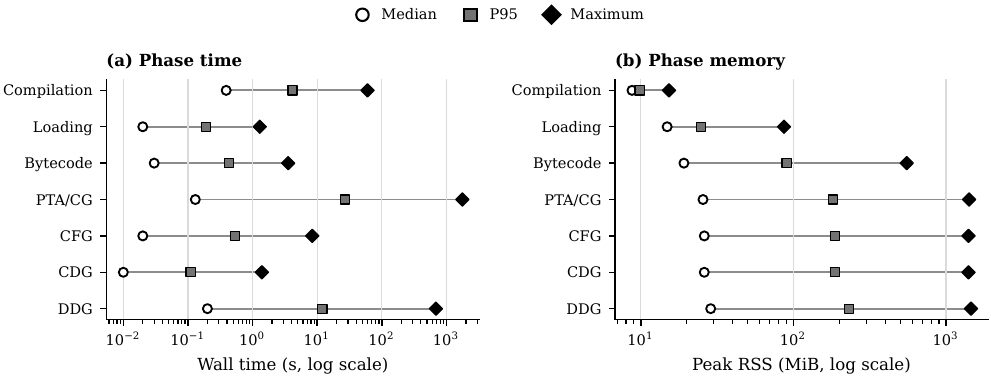}
\caption{Per-phase wall time and peak RSS. Markers show the median, P95, and
maximum on logarithmic axes; lines connect each phase's observed range.}
\Description{Two grayscale range plots compare median, 95th-percentile, and
maximum time and memory across compilation, loading, bytecode, PTA and call
graph, CFG, CDG, and DDG phases.}
\label{fig:rq4-component-costs}
\end{figure}

\noindent\textbf{Component cost and scaling.}
\Cref{fig:rq4-component-costs,tab:rq4-components} report wall time and process
RSS observed within each phase. The scale proxy is native instructions for
compilation, loading, and bytecode analysis; constraints for PTA/CG; blocks
plus edges for CFG and CDG; and nodes plus edges for DDG. The two $\rho$
columns give Spearman correlations between that proxy and phase time or RSS.
These are rank relationships, not linear-growth estimates.
All core analysis phases show strong time correlations (0.960--0.991) and
memory correlations (0.936--0.994) with their scale proxies. PTA/CG has the
largest upper-tail time: 27.03 seconds at P95 and 1,759.70 seconds at the
maximum. DDG follows at 12.02 and 686.23 seconds. Compilation memory is nearly
constant, which explains its lower memory correlation.

\begin{table}[H]
\centering
\caption{Per-phase cost and scaling. RSS is the process peak observed during
the phase. $\rho_t$ and $\rho_m$ correlate the phase's scale proxy with time
and RSS.}
\label{tab:rq4-components}
\resizebox{0.9\linewidth}{!}{%
\begin{tabular}{llrrrrrrrr}
\toprule
& & \multicolumn{3}{c}{Wall time (s)} & \multicolumn{3}{c}{Peak RSS (MiB)} & & \\
\cmidrule(lr){3-5}\cmidrule(lr){6-8}
Phase & Scale proxy & Median & P95 & Max. & Median & P95 & Max. & $\rho_t$ & $\rho_m$ \\
\midrule
Compilation & Instructions & 0.39 & 4.15 & 60.10 & 8.75 & 9.82 & 15.29 & 0.899 & 0.314 \\
Loading & Instructions & 0.02 & 0.19 & 1.29 & 14.87 & 24.88 & 86.70 & 0.951 & 0.917 \\
Bytecode & Instructions & 0.03 & 0.43 & 3.55 & 19.15 & 90.23 & 552.06 & 0.985 & 0.994 \\
PTA/CG & Constraints & 0.13 & 27.03 & 1,759.70 & 25.54 & 181.36 & 1,415.29 & 0.982 & 0.986 \\
CFG & Blocks+edges & 0.02 & 0.54 & 8.37 & 26.04 & 187.21 & 1,402.70 & 0.982 & 0.940 \\
CDG & Blocks+edges & 0.01 & 0.11 & 1.39 & 26.04 & 187.07 & 1,402.70 & 0.991 & 0.936 \\
DDG & Nodes+edges & 0.20 & 12.02 & 686.23 & 28.68 & 231.95 & 1,457.03 & 0.960 & 0.993 \\
\bottomrule
\end{tabular}%
}
\end{table}

\noindent\textbf{Answer to RQ4.}
Across real-package analyses (98.48\% of the corpus), \tech has a median end-to-end cost of
0.95 seconds and 28.40\,MiB, a P95 cost of 45.97 seconds and 231.95\,MiB, and
a maximum cost of 1,782.28 seconds and 1,457.03\,MiB. Both resources increase
with package size. PTA/CG and DDG account for most upper-tail analysis time.

\subsection{RQ5: Bundled-Bytecode Applicability}
\label{sec:rq5}

We select every package in the frozen real-package corpus containing a bundled
\texttt{.pyc} file. Thirteen of the 2,500 samples qualify, containing 951
\texttt{.pyc} files. We identify the generating runtime from the four-byte
magic header. As \cref{tab:rq5-population} shows, 732
files across 11 packages match CPython 3.10--3.14. The remaining files target
older CPython releases, PyPy, or an unknown untagged format.

\begin{table}[H]
\centering
\caption{Bundled-bytecode files by runtime family.}
\label{tab:rq5-population}
\scriptsize
\setlength{\tabcolsep}{6pt}
\resizebox{0.28\linewidth}{!}{%
\begin{tabular}{lr}
\toprule
Runtime family & Files \\
\midrule
CPython 3.10--3.14 & 732 \\
Older CPython      & 37 \\
PyPy               & 181 \\
Unknown/untagged   & 1 \\
\midrule
Total              & 951 \\
\bottomrule
\end{tabular}}
\end{table}

\noindent\textbf{Analysis completion.}
We pass each supported file directly to the matching \tech build. The analysis
uses the bundled bytecode directly. All 732 files complete bytecode recovery, PTA, CG,
CFG, CDG, and DDG construction. In total, \tech recovers 16,055 code objects.
The median file contains 8 code objects, the P95 contains 71, and the largest
contains 1,326.

\noindent\textbf{Cost.}
\Cref{tab:rq5-results} reports results by bytecode version. Across all files,
median and P95 wall time are 0.034 and 0.182 seconds; median and P95 peak RSS
are 15.27 and 19.50\,MiB. The maximum is 463.04 seconds and 866.95\,MiB for a
CPython 3.11 application module with 250 code objects and 27,799 native
instructions.

\begin{table}[H]
\centering
\caption{Direct bundled-bytecode analysis by CPython version. Every file
produces all five analysis products.}
\label{tab:rq5-results}
\resizebox{0.65\linewidth}{!}{%
\begin{tabular}{lrrrrrrr}
\toprule
& & \multicolumn{3}{c}{Wall time (s)} & \multicolumn{3}{c}{Peak RSS (MiB)} \\
\cmidrule(lr){3-5}\cmidrule(lr){6-8}
CPython & Files & Median & P95 & Max. & Median & P95 & Max. \\
\midrule
3.10 & 266 & 0.043 & 0.213 & 4.742 & 15.03 & 19.39 & 73.11 \\
3.11 & 209 & 0.033 & 0.182 & 463.044 & 14.76 & 19.56 & 866.95 \\
3.12 & 230 & 0.033 & 0.172 & 0.957 & 15.60 & 19.55 & 28.12 \\
3.13 & 13 & 0.034 & 0.085 & 0.119 & 15.46 & 17.56 & 18.14 \\
3.14 & 14 & 0.034 & 0.107 & 0.233 & 16.18 & 18.19 & 18.94 \\
\midrule
All & 732 & 0.034 & 0.182 & 463.044 & 15.27 & 19.50 & 866.95 \\
\bottomrule
\end{tabular}%
}
\end{table}

This study measures direct applicability to bundled bytecode in the frozen
corpus. The bundled-bytecode subset is reported separately from the full
efficiency corpus.

\noindent\textbf{Answer to RQ5.}
\tech directly analyzes all 732 bundled CPython 3.10--3.14 bytecode files and
constructs all five analysis products, recovering 16,055 code objects. The median
file takes 0.034 seconds and 15.27\,MiB.

\section{Discussion and Threats to Validity} \label{sec:diss}

\noindent\textbf{Summary of findings.}
The evaluation covers semantic accuracy, version behavior, design factors, and
real-package cost. On CPython 3.10, the default insensitive configuration
reaches 91.40\% candidate precision and 100\% recall on \bench; complete
sensitivity reaches 94.97\% candidate precision with the same recall. PTA
contributes 29 of the 54 false positives, and CG contributes two. On the
independent PyCG benchmark, PyCG and \tech both reach 111 complete and 103
sound call-graph cases. Across CPython 3.10--3.14, exact predictions agree on
99.55\% of version-invariant queries. RQ3 measures the effects of field
modeling, exception flow, sensitivity, and interprocedural DDG stitching. RQ4
completes 2,462 of 2,500 packages (98.48\%) with a median end-to-end cost of
0.95 seconds and 28.40\,MiB over successful runs. RQ5 directly analyzes all
732 bundled bytecode files for supported CPython versions.

The sensitivity experiment measures uniform applications of the Flow, Context,
and Path attributes. Explicit--All matches Complete. Sparse policies use the
same explicit function-level configuration described in
\cref{sec:design:sensitivity}.

\vspace{3pt}
\noindent\textbf{Modeled domain and dynamic Python behavior.}
The modeled domain determines the scope of the analysis~\cite{livshits2015soundiness}.
Reflection, dynamic code generation, native extensions, and external services
can introduce behavior beyond the modeled package domain. \tech records these cases
through typed unresolved groups and coverage summaries. Clients can project
these summaries onto their own finite candidate domains; for example, a
security client can use a broader projection than an optimization client.
For async programs, the CFG represents suspension and resume points. Scheduling
and interleavings require an event and scheduler model~\cite{python-asyncio}.

\noindent\textbf{Threats to validity.}

\begin{itemize}[leftmargin=1.5em,itemsep=2pt,topsep=3pt,parsep=0pt]
  \item \textbf{Construct validity.} Candidate precision and recall are measured
        over declared points-to memberships and finite graph paths. The
        candidate universe covers selected memberships and paths and is not an
        exhaustive graph oracle. Negative candidates are fault-revealing
        alternatives, so candidate precision measures rejection within this
        declared universe.

  \item \textbf{Internal validity.} Manual ground truth can contain mistakes.
        \bench uses small single-purpose programs, directly inspectable DOT/PTA
        artifacts, schema checks, and analyzer-independent generation of
        secondary structural facts. Analyzer output is not used to create the
        oracle. \tech and \bench were developed together, so implementation
        assumptions may influence candidate design. The independent PyCG
        benchmark supplies a separately developed oracle for the shared
        call-graph relation.

  \item \textbf{Ablation validity.} Each compile-time control removes one
        intended mechanism. Downstream fixed-point propagation can amplify a
        local change. The controls isolate field modeling, exception flow, and
        interprocedural DDG stitching; RQ1 evaluates the remaining modeled
        features through semantic candidates.

  \item \textbf{External validity.} Microbenchmarks isolate causes; production
        packages add scale, framework conventions, and reflective behavior. The
        frozen, size-stratified 2,500-package study measures scale and completion
        behavior. PyPI source distributions represent one part of the Python
        ecosystem. The version study covers CPython 3.10--3.14. Results for
        other CPython releases and other Python implementations require separate
        evaluation.
\end{itemize}

\section{Related Work} \label{sec:related}

\noindent\textbf{Static analysis of Python.}
PyCG constructs call graphs from Python source by propagating assignment
relations among identifiers and supports modules, closures, and
inheritance~\cite{salis2021pycg}. JARVIS adds application-centered, on-the-fly
call-graph construction with flow-sensitive type inference and strong
updates~\cite{huang2023jarvis}. These systems provide purpose-built Python call-graph analyses. \tech analyzes
native CPython code objects and provides CFG, CDG, PTA, CG, and DDG relations
over one operand-stack-aware analysis state.

Scalpel is a general source-level Python analysis framework that exposes CFG,
call-graph, import, and alias-analysis facilities~\cite{li2022scalpel}.
\texttt{codeanalyzer-python} provides source-level call, control, symbol, and dependence
relations through a reusable backend~\cite{canpy}.
\texttt{python\_graphs} constructs source-level control-flow, data-flow,
syntactic, and lexical program graphs, primarily for machine-learning
clients~\cite{bieber2022pythongraphs}. These projects provide reusable Python analysis infrastructure. \tech uses
native CPython code objects as its input boundary, adapts bytecode semantics by
CPython version, and couples PTA/CG facts with the other analysis products.

MOPSA derives a flow- and context-sensitive Python type analysis by abstract
interpretation of a concrete Python semantics and models objects, containers,
and exceptions~\cite{monat2020python}. Pytype interprets CPython bytecode
abstractly to infer and check types; its maintainers identify bytecode
instability as a challenge for version support~\cite{pytype}. MOPSA provides abstract-interpretation-based type analysis, and Pytype
provides bytecode-level type inference and checking. \tech produces coordinated
native CFG, CDG, PTA, CG, and DDG facts while retaining code-object identities
and native bytecode offsets across version adapters.

\vspace{3pt}
\noindent\textbf{Points-to and call-graph analysis.}
Andersen's inclusion-based analysis is the foundation for \tech's may-points-to
solver~\cite{andersen1994}. PoTo adapts Andersen-style analysis to Python
source, translates AST constructs to three-address code, and uses concrete
evaluation for external-library expressions~\cite{rakamnouykit2025poto}.
PoToCG applies those facts to call-graph construction. \tech applies inclusion
constraints to native operand-stack and object flow, couples those facts with
call-graph construction, and exposes explicit coverage for behavior beyond the modeled package domain.
Clients may additionally assign flow sensitivity, context sensitivity, and
bounded path partitioning to individual functions.
Python's first-class callables make PTA and CG mutually dependent. PyCG, JARVIS,
and PoToCG resolve related cycles through assignment, type, or points-to
information. On-the-fly call-graph construction is also established in
points-to analysis for object-oriented languages~\cite{grove1997callgraph,lhotak2003spark}.
In \tech, the coupled fixed point is the package-level prerequisite, and DDG
construction reuses its activation records.

\noindent\textbf{General analysis frameworks.}
Soot provides reusable analyses and intermediate representations for Java
bytecode~\cite{valleerai2010soot}. WALA has also been used as the analysis
framework for Ariadne's Python analysis~\cite{dolby2018ariadne}. SVF couples
pointer information with reusable value-flow graphs over LLVM
IR~\cite{sui2016svf}, while Doop expresses configurable Java points-to analyses
as declarative relations~\cite{bravenboer2009doop}. These frameworks organize multiple analyses around shared semantic
representations. \tech uses the same organization for native CPython code
objects and includes release-specific bytecode semantics, code-object
identities, and native bytecode offsets in the analysis contract.

\vspace{3pt}
\noindent\textbf{Bytecode and benchmarks.}
CPython's official documentation states that bytecode can change across
releases~\cite{python-dis}. A prior ecosystem study reports native bytecode in package artifacts,
including files without local source~\cite{chen2026beyondsource}. \tech
provides a version-aware analysis layer for this representation.
PyCG includes small semantic programs and real packages in its
evaluation~\cite{salis2021pycg}. DyPyBench provides 50 executable Python
projects and infrastructure for dynamic analyses, including a static/dynamic
call-graph comparison~\cite{bouzenia2024dypybench}. TypeEvalPy organizes small
Python programs by type-system feature and uses a canonical result
format~\cite{venkatesh2024typeevalpy}. PointerBench uses manually validated
microprograms to expose points-to-analysis differences~\cite{prakash2019pointeval}.
\bench provides explicit expected and negative PTA memberships and CG, CFG,
CDG, and DDG paths for small packages. Its canonical result schema lets analyzers
compare semantic facts without sharing internal nodes.

\section{Conclusion}\label{sec:conclude}

\tech provides an integrated foundation for package-level analysis of native
CPython bytecode. Version adapters isolate changing opcode, stack, call, and
exception semantics while preserving code-object identities and native bytecode
offsets. An operand-stack-aware Andersen PTA and call graph grow together to a fixed point, and
their activation records support exception-aware CFG, block-level CDG, and
interprocedural DDG clients. Function-level sensitivity refines selected PTA
activations, and coverage summaries expose unresolved behavior through the same
API.

On 201 \bench cases, the CPython 3.10 default configuration reaches 91.40\%
candidate precision and 100\% recall across 1,733 fixed semantic candidates;
complete sensitivity reaches 94.97\% candidate precision with the same recall.
On PyCG's independent benchmark, \tech matches PyCG's 111 complete and 103
sound cases. Across CPython 3.10--3.14, 1,328 of 1,334 version-invariant
queries receive identical predictions. The real-package study completes 2,462
of 2,500 packages, and all 732 supported bundled \texttt{.pyc} files produce all five analysis
products. The framework and machine-readable ground truth provide a concrete reference
for bytecode-level Python analysis.

\bibliographystyle{ACM-Reference-Format}
\bibliography{reference}

@phdthesis{andersen1994,
  author={Andersen, Lars Ole},
  title={Program Analysis and Specialization for the C Programming Language},
  number={DIKU Report 94/19},
  year={1994},
  school={University of Copenhagen}
}

@inproceedings{cousot1977abstract,
  author={Cousot, Patrick and Cousot, Radhia},
  title={Abstract Interpretation: A Unified Lattice Model for Static
               Analysis of Programs by Construction or Approximation of
               Fixpoints},
  booktitle={Proceedings of the 4th ACM SIGACT-SIGPLAN Symposium on
               Principles of Programming Languages},
  publisher={ACM},
  pages={238--252},
  year={1977},
  doi={10.1145/512950.512973},
  address={New York, NY, USA}
}

@inproceedings{grove1997callgraph,
  author={Grove, David and DeFouw, Greg and Dean, Jeffrey and Chambers, Craig},
  title={Call Graph Construction in Object-Oriented Languages},
  booktitle={Proceedings of the 12th ACM SIGPLAN Conference on
               Object-Oriented Programming, Systems, Languages, and
               Applications},
  publisher={ACM},
  pages={108--124},
  year={1997},
  doi={10.1145/263698.264352},
  address={New York, NY, USA}
}

@inproceedings{lhotak2003spark,
  author={Lhot\'{a}k, Ond\v{r}ej and Hendren, Laurie J.},
  title={Scaling Java Points-to Analysis Using {Spark}},
  booktitle={Compiler Construction},
  publisher={Springer},
  volume={2622},
  pages={153--169},
  year={2003},
  doi={10.1007/3-540-36579-6_12},
  address={Berlin, Heidelberg},
  series={Lecture Notes in Computer Science}
}

@inproceedings{mauborgne2005trace,
  author={Mauborgne, Laurent and Rival, Xavier},
  title={Trace Partitioning in Abstract Interpretation Based Static
               Analyzers},
  booktitle={Programming Languages and Systems: 14th European Symposium on
               Programming},
  publisher={Springer},
  volume={3444},
  pages={5--20},
  year={2005},
  doi={10.1007/978-3-540-31987-0_2},
  address={Berlin, Heidelberg},
  series={Lecture Notes in Computer Science}
}

@article{livshits2015soundiness,
  author={Livshits, Benjamin and Sridharan, Manu and Smaragdakis, Yannis and Lhot\'{a}k, Ond\v{r}ej and Amaral, J. Nelson and Chang, Bor-Yuh Evan and Guyer, Samuel Z. and Khedker, Uday P. and M\o{}ller, Anders and Vardoulakis, Dimitrios},
  title={In Defense of Soundiness: A Manifesto},
  journal={Communications of the ACM},
  publisher={ACM},
  volume={58},
  number={2},
  pages={44--46},
  year={2015},
  doi={10.1145/2644805}
}

@inproceedings{salis2021pycg,
  author={Salis, Vitalis and Sotiropoulos, Thodoris and Louridas, Panos and Spinellis, Diomidis and Mitropoulos, Dimitris},
  title={{PyCG}: Practical Call Graph Generation in Python},
  booktitle={Proceedings of the 43rd IEEE/ACM International Conference on
               Software Engineering},
  publisher={IEEE},
  pages={1646--1657},
  year={2021},
  doi={10.1109/ICSE43902.2021.00146},
  address={Los Alamitos, CA, USA}
}

@article{li2022scalpel,
  author={Li, Li and Wang, Jiawei and Quan, Haowei},
  title={Scalpel: The Python Static Analysis Framework},
  journal={CoRR},
  volume={abs/2202.11840},
  year={2022},
  eprint={2202.11840},
  archiveprefix={arXiv},
  url={https://arxiv.org/abs/2202.11840}
}

@inproceedings{rakamnouykit2025poto,
  author={Rak-amnouykit, Ingkarat and Milanova, Ana and Baudart, Guillaume and Hirzel, Martin and Dolby, Julian},
  title={{PoTo}: A Hybrid Andersen's Points-To Analysis for Python},
  booktitle={39th European Conference on Object-Oriented Programming},
  publisher={Schloss Dagstuhl--Leibniz-Zentrum f\"{u}r Informatik},
  volume={333},
  pages={27:1--27:29},
  year={2025},
  doi={10.4230/LIPIcs.ECOOP.2025.27},
  address={Dagstuhl, Germany},
  series={Leibniz International Proceedings in Informatics}
}

@article{huang2023jarvis,
  author={Huang, Kaifeng and Yan, Yixuan and Chen, Bihuan and Tao, Zixin and Peng, Xin},
  title={Scalable and Precise Application-Centered Call Graph Construction
             for Python},
  journal={CoRR},
  volume={abs/2305.05949},
  year={2023},
  eprint={2305.05949},
  archiveprefix={arXiv},
  url={https://arxiv.org/abs/2305.05949}
}

@article{bieber2022pythongraphs,
  author={Bieber, David and Shi, Kensen and Maniatis, Petros and Sutton, Charles and Hellendoorn, Vincent and Johnson, Daniel and Tarlow, Daniel},
  title={A Library for Representing Python Programs as Graphs for Machine
             Learning},
  journal={CoRR},
  volume={abs/2208.07461},
  year={2022},
  eprint={2208.07461},
  archiveprefix={arXiv},
  url={https://arxiv.org/abs/2208.07461}
}

@article{bouzenia2024dypybench,
  author={Bouzenia, Islem and Krishan, Bajaj Piyush and Pradel, Michael},
  title={{DyPyBench}: A Benchmark of Executable Python Software},
  journal={Proceedings of the ACM on Software Engineering},
  publisher={ACM},
  volume={1},
  number={FSE},
  pages={338--358},
  year={2024},
  doi={10.1145/3643742}
}

@inproceedings{venkatesh2024typeevalpy,
  author={Venkatesh, Ashwin Prasad S. and Sabu, Samkutty and Wang, Jiawei and Mir, Amir M. and Li, Li and Bodden, Eric},
  title={{TypeEvalPy}: A Micro-benchmarking Framework for Python Type
               Inference Tools},
  booktitle={Companion Proceedings of the 46th International Conference on
               Software Engineering},
  publisher={ACM},
  pages={49--53},
  year={2024},
  doi={10.1145/3639478.3640033},
  address={New York, NY, USA}
}

@article{prakash2019pointeval,
  author={Prakash, Jyoti and Tiwari, Abhishek and Hammer, Christian},
  title={{PointEval}: On the Impact of Pointer Analysis Frameworks},
  journal={CoRR},
  volume={abs/1912.00429},
  year={2019},
  eprint={1912.00429},
  archiveprefix={arXiv},
  url={https://arxiv.org/abs/1912.00429}
}

@inproceedings{monat2020python,
  author={Monat, Rapha\"{e}l and Ouadjaout, Abdelraouf and Min\'{e}, Antoine},
  title={Static Type Analysis by Abstract Interpretation of Python
               Programs},
  booktitle={34th European Conference on Object-Oriented Programming},
  publisher={Schloss Dagstuhl--Leibniz-Zentrum f\"{u}r Informatik},
  volume={166},
  pages={17:1--17:29},
  year={2020},
  doi={10.4230/LIPIcs.ECOOP.2020.17},
  address={Dagstuhl, Germany},
  series={Leibniz International Proceedings in Informatics}
}

@article{chen2026beyondsource,
  author={Chen, Baihong and Xie, Tian and Li, Wen},
  title={Beyond Source: An Empirical Study of Python Bytecode Security
             Risks},
  journal={CoRR},
  volume={abs/2608.12853},
  year={2026},
  eprint={2608.12853},
  archiveprefix={arXiv},
  url={https://arxiv.org/abs/2608.12853}
}

@misc{python-dis,
  author={{Python Software Foundation}},
  title={\texttt{dis}---Disassembler for Python Bytecode},
  year={2026},
  note={Accessed September 15, 2026},
  howpublished={\url{https://docs.python.org/3/library/dis.html}}
}

@misc{pep659,
  author={Shannon, Mark},
  title={{PEP 659}---Specializing Adaptive Interpreter},
  year={2021},
  note={Implemented in Python 3.11},
  howpublished={\url{https://peps.python.org/pep-0659/}}
}

@misc{pep552,
  author={Peterson, Benjamin},
  title={{PEP 552}---Deterministic \texttt{pyc}s},
  year={2017},
  howpublished={\url{https://peps.python.org/pep-0552/}}
}

@misc{python-importlib,
  author={{Python Software Foundation}},
  title={\texttt{importlib}---The Implementation of \texttt{import}},
  year={2026},
  note={Accessed September 17, 2026},
  howpublished={\url{https://docs.python.org/3/library/importlib.html\#importlib.machinery.SourcelessFileLoader}}
}

@misc{python-marshal,
  author={{Python Software Foundation}},
  title={\texttt{marshal}---Internal Python Object Serialization},
  year={2026},
  note={Accessed September 17, 2026},
  howpublished={\url{https://docs.python.org/3/library/marshal.html}}
}

@misc{python-asyncio,
  author={{Python Software Foundation}},
  title={Coroutines and Tasks},
  year={2026},
  note={Accessed September 17, 2026},
  howpublished={\url{https://docs.python.org/3/library/asyncio-task.html}}
}

@misc{pep709,
  author={Meyer, Carl},
  title={{PEP 709}---Inlined Comprehensions},
  year={2023},
  note={Python 3.12},
  howpublished={\url{https://peps.python.org/pep-0709/}}
}

@misc{python-datamodel,
  author={{Python Software Foundation}},
  title={The Python Language Reference: Data Model},
  year={2026},
  note={Accessed September 15, 2026},
  howpublished={\url{https://docs.python.org/3/reference/datamodel.html}}
}

@misc{python-execution-model,
  author={{Python Software Foundation}},
  title={The Python Language Reference: Execution Model},
  year={2026},
  note={Accessed September 17, 2026},
  howpublished={\url{https://docs.python.org/3/reference/executionmodel.html}}
}

@misc{python-import-system,
  author={{Python Software Foundation}},
  title={The Python Language Reference: The Import System},
  year={2026},
  note={Accessed September 17, 2026},
  howpublished={\url{https://docs.python.org/3/reference/import.html}}
}

@incollection{sharir1981,
  author={Sharir, Micha and Pnueli, Amir},
  editor={Steven S. Muchnick and Neil D. Jones},
  title={Two Approaches to Interprocedural Data Flow Analysis},
  booktitle={Program Flow Analysis: Theory and Applications},
  publisher={Prentice-Hall},
  pages={189--234},
  year={1981},
  address={Englewood Cliffs, NJ}
}

@article{ferrante1987pdg,
  author={Ferrante, Jeanne and Ottenstein, Karl J. and Warren, Joe D.},
  title={The Program Dependence Graph and Its Use in Optimization},
  journal={ACM Transactions on Programming Languages and Systems},
  publisher={ACM},
  volume={9},
  number={3},
  pages={319--349},
  year={1987},
  doi={10.1145/24039.24041}
}

@inproceedings{kildall1973,
  author={Kildall, Gary A.},
  title={A Unified Approach to Global Program Optimization},
  booktitle={Proceedings of the 1st ACM SIGACT-SIGPLAN Symposium on
               Principles of Programming Languages},
  publisher={ACM},
  pages={194--206},
  year={1973},
  doi={10.1145/512927.512945},
  address={New York, NY, USA}
}

@misc{pypi-bigquery,
  author={{Python Package Index}},
  title={{PyPI} Docs: BigQuery Datasets},
  year={2026},
  note={Accessed September 17, 2026},
  howpublished={\url{https://docs.pypi.org/api/bigquery/}}
}

@misc{python-core-metadata,
  author={{Python Packaging Authority}},
  title={Core Metadata Specifications},
  year={2026},
  note={Accessed September 17, 2026},
  howpublished={\url{https://packaging.python.org/en/latest/specifications/core-metadata/}}
}

@misc{canpy,
  author={{CodeLLM DevKit Developers}},
  title={\texttt{codeanalyzer-python}: Python Static Analysis Backend
                  for CLDK},
  year={2026},
  note={Accessed September 15, 2026},
  howpublished={\url{https://github.com/codellm-devkit/codeanalyzer-python}}
}

@misc{pytype,
  author={{Google}},
  title={\texttt{pytype}: A Static Type Analyzer for Python Code},
  year={2026},
  note={Archived September 3, 2026; accessed September 17, 2026},
  howpublished={\url{https://github.com/google/pytype}}
}

@inproceedings{valleerai2010soot,
  author={Vall\'{e}e-Rai, Raja and Co, Phong and Gagnon, Etienne and Hendren, Laurie J. and Lam, Patrick and Sundaresan, Vijay},
  title={Soot: A Java Bytecode Optimization Framework},
  booktitle={CASCON First Decade High Impact Papers},
  publisher={ACM},
  pages={214--224},
  year={2010},
  doi={10.1145/1925805.1925818}
}

@inproceedings{sui2016svf,
  author={Sui, Yulei and Xue, Jingling},
  title={{SVF}: Interprocedural Static Value-Flow Analysis in {LLVM}},
  booktitle={Proceedings of the 25th International Conference on Compiler Construction},
  publisher={ACM},
  pages={265--266},
  year={2016},
  doi={10.1145/2892208.2892235}
}

@inproceedings{bravenboer2009doop,
  author={Bravenboer, Martin and Smaragdakis, Yannis},
  title={Strictly Declarative Specification of Sophisticated Points-to Analyses},
  booktitle={Proceedings of the 24th ACM SIGPLAN Conference on Object-Oriented Programming, Systems, Languages, and Applications},
  publisher={ACM},
  pages={243--262},
  year={2009},
  doi={10.1145/1640089.1640108}
}

@article{dolby2018ariadne,
  author={Dolby, Julian and Shinnar, Avraham and Allain, Allison and Reinen, Jenna},
  title={Ariadne: Analysis for Machine Learning Programs},
  journal={CoRR},
  volume={abs/1805.04058},
  year={2018},
  eprint={1805.04058},
  archiveprefix={arXiv},
  url={https://arxiv.org/abs/1805.04058}
}

\newpage
\appendixcontent
\section{Version-Adapter Reference}\label{app:version-reference}

The main design exposes semantic operations to the shared analyses, while the
version adapters handle raw opcode numbers. This appendix records the
release-specific work behind that boundary. The rows in \cref{tab:running-version-deltas} are generated by
compiling the program in \cref{fig:motivating} with each supported
interpreter. The rows show how the same source constructs map to release-specific native
instruction sequences~\cite{python-dis}.

\begin{table}[H]
\centering
\scriptsize
\setlength{\tabcolsep}{3pt}
\caption{Native lowerings of the running example.}
\label{tab:running-version-deltas}
\resizebox{\linewidth}{!}{%
\begin{tabular}{p{0.07\linewidth}p{0.24\linewidth}p{0.34\linewidth}p{0.26\linewidth}}
\toprule
Version & Context manager & Calls and captured closure & Protected regions \\
\midrule
3.10 & \texttt{SETUP\_WITH} &
\texttt{CALL\_FUNCTION}; \texttt{LOAD\_CLOSURE} and flag-bearing
\texttt{MAKE\_FUNCTION} & Reconstructed from \texttt{SETUP\_*} block
instructions \\
3.11 & \texttt{BEFORE\_WITH} &
\texttt{PRECALL}+\texttt{CALL}; \texttt{MAKE\_CELL},
\texttt{LOAD\_CLOSURE}, and \texttt{MAKE\_FUNCTION} & Decoded from
\texttt{co\_exceptiontable} \\
3.12 & \texttt{BEFORE\_WITH} &
\texttt{CALL}; the same explicit cell operations as 3.11 & Decoded from
\texttt{co\_exceptiontable} \\
3.13 & \texttt{BEFORE\_WITH} &
\texttt{CALL}; \texttt{LOAD\_FAST}, \texttt{MAKE\_FUNCTION}, then
\texttt{SET\_FUNCTION\_ATTRIBUTE} for the closure & Decoded from
\texttt{co\_exceptiontable} \\
3.14 & Two \texttt{LOAD\_SPECIAL} operations followed by \texttt{CALL} &
\texttt{CALL}; \texttt{LOAD\_FAST\_BORROW}, \texttt{MAKE\_FUNCTION}, then
\texttt{SET\_FUNCTION\_ATTRIBUTE} & Decoded from
\texttt{co\_exceptiontable} \\
\bottomrule
\end{tabular}
}%
\end{table}

For the 3.12 trace, \texttt{BEFORE\_WITH} at offset 64 becomes
\textsc{EnterContext} with one stack input and two outputs. The protected
range printed as offsets 66--84 becomes the half-open interval $[66,86)$.
For 3.10, the adapter reconstructs the same semantic region from block-setup
instructions. For 3.14, separate special-method loads preserve the
\texttt{ContextEnter} and \texttt{ContextExit} protocol identities. The
shared core consumes native origins and the semantic records in
\cref{tab:semantic-record}; release-specific opcode names stay in the adapter.

\begin{table}[H]
\centering
\footnotesize
\setlength{\tabcolsep}{5pt}
\renewcommand{\arraystretch}{1.08}
\caption{Version-neutral adapter records with running-example values.}
\label{tab:semantic-record}
\resizebox{\linewidth}{!}{%
\begin{tabular}{@{}>{\raggedright\arraybackslash}p{0.17\linewidth}
                    >{\raggedright\arraybackslash}p{0.38\linewidth}
                    >{\raggedright\arraybackslash}p{0.34\linewidth}@{}}
\toprule
Record & Meaning and consumer & Example from \cref{fig:motivating} \\
\midrule
Origin & Code-object ID and native bytecode offset; used by every graph and report. &
$\langle id_{run},64\rangle$ identifies the context-entry instruction. \\
Operation & Shared operation and decoded operand; used by PTA, CG, and DDG. &
\mbox{\textsc{EnterContext}} at offset 64 and
\mbox{\textsc{StoreLocal}}($value$) at 66. \\
Control & Absolute target and outcome polarity; used by CFG and CDG. &
The false outcome at offset 70 targets offset 106. \\
Stack & Normal and jump-edge inputs, outputs, and peeks; used by CFG, PTA, and
DDG. & \mbox{\textsc{EnterContext}} consumes one value and produces two. \\
Call & Arity, receiver, keyword layout, and expanded arguments; used by PTA and
CG. & The call at offset 10 passes one value to the captured callback. \\
Lexical access & Fast-local, cell-value, cell-reference, or cell-creation kind;
used by PTA and DDG. & \texttt{LOAD\_DEREF callback} reads the captured cell's
contents. \\
Protocol & Implicit-feature and compatible-method IDs; used by PTA, CG, and the coverage summary. & Context entry and exit identify
\texttt{Guard.\_\_enter\_\_} and \texttt{Guard.\_\_exit\_\_}. \\
Exception region & Protected range, handler, restored depth, exception items,
and \texttt{lasti}; used by CFG and DDG, with CDG derived from the resulting CFG. &
$\langle66,86,140,1,2,\mathsf{true}\rangle$. \\
\bottomrule
\end{tabular}
}%
\end{table}

\noindent\textbf{Reading one record.}
For \texttt{BEFORE\_WITH} at offset 64, the adapter emits the origin,
\textsc{EnterContext} operation, one-to-two stack transfer, and context-entry
and context-exit protocol IDs shown across the table. The CFG consumes its control and stack fields; PTA and CG consume its protocol
fields; reports keep the native origin. The raw \texttt{BEFORE\_WITH} opcode
remains confined to the version adapter.

\noindent\textbf{Transfer rules.}
For a normal edge, the common stack transfer is
$d'=d-i+o$, where $i$ and $o$ are the decoded input and output counts.
Conditional instructions carry a second pair $(i_j,o_j)$ for the jump edge.
An exception region $\langle s,e,h,d,x,\ell\rangle$ discards the transient
suffix at a protected instruction, restores depth $d$, adds $x$ handler
items, and transfers to $h$. This representation covers changes in opcode
numbers, jump bases, inline caches, call slots, and exception encodings, so
graph builders use the same transfer rules across versions.

\section{Python-Feature Modeling Reference}
\label{app:python-feature-reference}

Version adapters identify operations; the shared Python layer assigns their
object and call semantics. \Cref{tab:python-feature-rules} summarizes the
rules used by the implementation. All stored entities use numeric
IDs. The names shown here are presentation labels decoded by reports. The
rules follow the Python execution, import, data, and task models~\cite{python-execution-model,
python-import-system,python-datamodel,python-asyncio}.

\begin{table}[H]
\centering
\scriptsize
\setlength{\tabcolsep}{4pt}
\caption{Shared rules for Python-specific behavior.}
\label{tab:python-feature-rules}
\resizebox{\linewidth}{!}{%
\begin{tabular}{p{0.18\linewidth}p{0.33\linewidth}p{0.38\linewidth}}
\toprule
Feature & Shared model & Coverage boundary \\
\midrule
Imports and aliases & Module bodies are roots. Module objects own exported
fields; relative levels, from-lists, and imported attributes preserve aliases. &
Computed names and external modules use a typed import group. \\
Functions and classes & Function objects point to code objects and attached
defaults or closure data. Class calls allocate instance objects and connect
visible class fields. & Dynamic metaclass and external constructor behavior
can remain typed and unresolved. \\
Captured variables & Cell creation, cell references, and cell-value access map
to an explicit cell object with a distinguished contents field. & Insensitive
activations may merge cells; selective context sensitivity separates them. \\
Attributes and methods & PTA restricts receiver objects. Instance function
loads create bound-method objects that retain function and receiver. & Dynamic
replacement and unproved descriptor behavior remain protocol alternatives. \\
Descriptors and properties & The model orders \texttt{\_\_getattribute\_\_},
descriptor access, and possible \texttt{\_\_getattr\_\_}; property calls use
the resolved receiver. & An unknown receiver retains the compatible method
set. \\
Operators and comparisons & Direct, reflected, and in-place methods are
selected in Python order. A possible \texttt{NotImplemented} return retains
the next alternative. & External operands or method bodies retain a compact
method-family group. \\
Truth and iteration & Truth tests select \texttt{\_\_bool\_\_} before
\texttt{\_\_len\_\_}. Iteration links iterator and next-method dispatch. &
Unknown receiver types retain only the compatible truth or iteration family. \\
Context managers & Entry and exit methods are recovered from the manager.
The entry result and retained cleanup callable follow the stack and exception
region. & Runtime method replacement remains a typed context group. \\
Containers and fields & Loads and stores use PTA-resolved base objects and
interned field IDs; collection elements use a summary element field. & Unknown
bases widen only the compatible heap facts. \\
Exceptions & Normalized protected regions create handler edges and restore the
handler stack. DDG propagation follows both normal and exceptional CFG edges. &
Exceptions raised only by unmodeled native behavior follow the coverage summary. \\
Coroutines & Suspension, resume, and stack effects are retained structurally. &
Scheduling and task interleavings require a separate asynchronous model. \\
\bottomrule
\end{tabular}
}%
\end{table}

\noindent\textbf{Motivating-example trace.}
In the running example, the class call creates $o_G$, context entry resolves
\texttt{Guard.\_\_enter\_\_}, and each \texttt{bind} activation attaches one
callback cell to an \texttt{invoke} closure. Callback loads reveal new targets.
The solver repeats until points-to sets and call edges stop growing; all graph
products reuse its objects, activations, code-object identities, and native
bytecode offsets.

\begin{figure}[H]
\centering
\begin{minipage}{0.6\linewidth}
\begin{tcolorbox}[codepiece,title={\sffamily\bfseries Imports, objects, protocols, and exceptions},
  fonttitle=\sffamily\small,colbacktitle=black!11,coltitle=black]
\begin{lstlisting}[language=Python,numbers=left,
  numberstyle=\tiny\color{codecomment},numbersep=4pt,
  xleftmargin=1.5em,basicstyle=\ttfamily\footnotesize]
from .model import Item as I

class Box:
    def __init__(self, item):
        self._item = item
    @property
    def item(self):
        return self._item
    def first_sum(self, values):
        try:
            for value in values:
                if value:
                    return self.item + value
        except ValueError:
            return I(0)
def use(seed, values):
    return Box(seed).first_sum(values)
\end{lstlisting}
\end{tcolorbox}
\end{minipage}
\caption{A companion example for Python-specific semantic rules.}
\Description{A short Python example combines an import alias, a class and
field, a property, a bound-method call, iteration, a truth test, operator
dispatch, and an exception handler.}
\label{fig:python-feature-example}
\end{figure}

\noindent\textbf{Additional feature example.}
The compact program in \cref{fig:python-feature-example} uses ordinary Python
syntax to exercise the other common rules.
Its version-specific instructions are normalized before these facts are
created:
\begin{itemize}[leftmargin=1.5em,itemsep=1pt,topsep=2pt,parsep=0pt]
  \item \textbf{Import and allocation.} The local name \texttt{I} points to
        the exported \texttt{Item} object. Calling \texttt{Box} allocates
        $o_B$, activates \texttt{Box.\_\_init\_\_}, and makes the
        \texttt{self} parameter point to $o_B$.
  \item \textbf{Fields, descriptors, and methods.} The store to
        \texttt{self.\_item} and the later load use the same interned field.
        Loading \texttt{item} invokes the package-defined property getter.
        Loading \texttt{first\_sum} from $o_B$ creates a bound method that
        retains both $o_B$ and its function.
  \item \textbf{Iteration and truth.} Normalized iteration operations select
        \texttt{\_\_iter\_\_} and \texttt{\_\_next\_\_}. The conditional
        selects \texttt{\_\_bool\_\_}, with \texttt{\_\_len\_\_} as the
        ordered fallback, and gives the CFG and CDG explicit branch outcomes.
  \item \textbf{Operator dispatch.} The addition records the direct
        \texttt{\_\_add\_\_} candidate and its reflected
        \texttt{\_\_radd\_\_} alternative. Resolved receiver objects narrow
        these candidates; any remaining compatible family is a typed group.
  \item \textbf{Exceptions.} The normalized protected region connects
        may-raise instructions in the loop to the matching handler. The
        handler call targets the same \texttt{Item} object named by alias
        \texttt{I}; DDG propagation follows both normal and exceptional edges.
\end{itemize}

\section{Selective-Sensitivity Configuration}\label{app:sensitivity}

Selective sensitivity lets a client refine PTA for selected functions while
the rest of the package remains insensitive. Each configuration entry maps a
numeric code-object ID to one or more sensitivity attributes: flow, context,
or bounded path partitioning. Context sensitivity distinguishes activations,
flow sensitivity distinguishes local state at program points, and path
partitioning separates bounded acyclic branch alternatives. Path partitioning
also enables flow sensitivity.
The following configuration applies context sensitivity to \texttt{bind} and
\texttt{invoke} in the running example. This separates the closure cells
created for the \texttt{left} and \texttt{right} callbacks. It also applies
bounded path partitioning to \texttt{run}, which distinguishes its two
acyclic branch outcomes. Unlisted functions keep the default insensitive
abstraction.

\begin{tcolorbox}[
  codepiece,
  breakable,
  title={\sffamily\bfseries Function-level PTA sensitivity configuration},
  fonttitle=\sffamily\small,
  colbacktitle=black!11,
  coltitle=black]
\begin{lstlisting}[language=C++,basicstyle=\ttfamily\footnotesize]
cpygraph::PTASensitivityConfiguration policy;
policy.level = cpygraph::PTASensitivityLevel::Selective;
policy.maximum_path_variants =
    cpygraph::kDefaultMaximumPTAPathVariants;
policy.functions = {
    {bind_id, cpygraph::PTASensitivity::Context},
    {invoke_id, cpygraph::PTASensitivity::Context},
    {run_id, cpygraph::PTASensitivity::Path},
};

auto graphs = cpygraph::package::GraphBuilder{}.analyze(
    package, {}, policy);
\end{lstlisting}

\tcblower
\small
\noindent\textbf{Function keys.}
\texttt{bind\_id}, \texttt{invoke\_id}, and \texttt{run\_id} are numeric
\texttt{CodeObjectId} values from the loaded package. Code-object IDs
distinguish nested functions that may share the same source-level name.

\vspace{2pt}
\noindent\textbf{Selected attributes.}
\texttt{Context} gives \texttt{bind} and \texttt{invoke} a bounded
two-call-site context. The two calls to \texttt{bind} therefore create
different closure-cell abstractions for \texttt{left} and \texttt{right}.
\texttt{Path} gives \texttt{run} bounded acyclic branch variants and
implicitly enables \texttt{Flow}. Attributes can be combined with the
bitwise-or operator. Every unlisted function remains insensitive.

\vspace{2pt}
\noindent\textbf{Path bound.}
\texttt{maximum\_path\_variants} limits the number of variants created for
each function. When the bound is reached, additional acyclic alternatives are
merged. Cyclic alternatives also merge, keeping the analysis state finite.

\vspace{2pt}
\noindent\textbf{Analysis modes.}
The default policy is \texttt{Insensitive}. \texttt{Selective} applies only
the listed function attributes. \texttt{Complete} applies flow, context, and
path sensitivity to every function. The same policy is accepted by the
combined package analysis and the individual graph entry points. PTA and CG
consume the policy during their shared fixed point, and the other analysis
products reuse the resulting package state.
\end{tcolorbox}

\end{document}